\documentclass{acmtog}

\acmVolume{VV}
\acmNumber{N}
\acmYear{YYYY}
\acmMonth{Month}
\acmArticleNum{XXX}
\acmdoi{10.1145/XXXXXXX.YYYYYYY}

\usepackage{subfigure}
\usepackage{color}
\usepackage{soul}
\usepackage[labelfont=bf,textfont=it]{caption}
\usepackage{wrapfig}

\begin{document}

\markboth{D.-M. Yan and P. Wonka}{Gap Processing for Adaptive Maximal Poisson-Disk Sampling}

\title{Gap Processing for Adaptive Maximal Poisson-Disk Sampling} 

\author{DONG-MING YAN
\affil{King Abdullah University of Science and Technology}
\and
PETER WONKA
\affil{King Abdullah University of Science and Technology}}




\acmformat{Yan, D.-M., and Wonka, P.. 2013. Gap Processing for Adaptive Maximal Poisson-Disk Sampling.  {ACM Trans. Graph.} 28, 4, Article xxx (xxx 2013), 15 pages.\newline  DOI $=$
xxx\newline http://doi.acm.org/xxx}

\def \wrt {w.r.t.~}
\def \wps {\mathcal{P}_{w}}
\def \rrt {\mathcal{RRT}}
\def \rt {\mathcal{RT}}
\def \pd {\mathcal{PD}}
\def \rpd {\mathcal{RPD}}
\def \vd {\mathcal{VD}}
\def \GT {\mathcal{GT}}
\def \gap{\mathcal{G}}
\def \mx {\mathbf{x}}
\def \mp {\mathbf{p}}
\def \mS {\mathbf{S}}
\def \mv {\mathbf{v}}
\def \mV {\mathbf{V}}
\def \mX {\mathbf{X}}
\def \mc {\mathbf{c}}
\def \cT {\mathcal{T}}
\def \cD {\mathcal{D}}
\def \cX {\mathcal{X}}
\def \bn {\mathcal{BN}}
\newcommand{\red}[1]{\textcolor[rgb]{1.00,0.00,0.00}{#1}}
\newcommand{\ndim}[1]{${#1}$-D}
\def \highdim{1}

\maketitle

\begin{bottomstuff}
This research was partially funded by National Natural Science Foundation of China (No. 61271431, 61172104 and 61271430), and the National Science Foundation.

Authors' addresses: D.-M. Yan (corresponding author), KAUST, email: yandongming@gmail.com; P. Wonka (corresponding author), KAUST, email: pwonka@gmail.com.
\end{bottomstuff}

\begin{abstract}
In this paper, we study the generation of maximal Poisson-disk sets with varying radii.
First, we present a geometric analysis of gaps in such disk sets. This analysis is the basis for maximal and adaptive sampling in Euclidean space and on manifolds. Second, we propose efficient algorithms and data structures to detect gaps and update gaps when disks are inserted, deleted, moved, or when their radii are changed. We build on the concepts of regular triangulations and the power diagram. Third, we show how our analysis contributes to the state of the art in surface remeshing.
\end{abstract}

\section{Introduction}
Maximal Poisson-disk sampling (MPS) can generate point sets with interesting properties. One issue in MPS is how to locate and approximate the gaps in an already sampled disk set.

In this paper, we study the geometry of gaps in disk sets. Given a set of disks with varying radii in a compact domain $\Omega$ in Euclidean space or on a manifold, we would like to know if they fully cover the domain or if there is a gap, i.e., if there is uncovered space to insert other disks into the disk set. We are interested in knowing if gaps exist, where the gaps are, and how to implement efficient gap processing operations.

There are three important papers that conduct such an analysis in the context of uniform 2D MPS: Dunbar and Humphreys~\shortcite{Dunbar2006}, Jones~\shortcite{Jones2006}, and Ebeida et al.~\shortcite{Ebeida2011}. In the first part of this paper, we present a simpler and more natural analysis that additionally extends to 1) disk sets with varying radii, 2) arbitrary dimensions, and 3) 2-manifolds. The main idea of our approach is to study gaps in the context of the regular triangulation and the power diagram~\cite{Aurenhammer1991} of the disk sets.

Besides the general curiosity about an interesting geometric problem, gaps in disk sets play an important role in sampling applications, e.g., sample generation for ray tracing, image stippling, video stippling, environment map sampling, surface remeshing, plant ecosystem simulation, texture synthesis, video synthesis, and particle-based simulation. By analyzing this set of interesting applications to find commonalities, we identified several operations that need to be performed efficiently: gap detection, gap clustering, gap primitive extraction, updating gap primitives when points are deleted or inserted, updating gap primitives when points move, and updating gap primitives when the sampling radius changes. The second part of the paper will introduce algorithms for these operations, based on the analysis in the first part.

While our algorithms do not improve all aforementioned applications, in the third part of this paper, we investigate an application to surface remeshing (as well as 2D meshing), where we can successfully improve the state of the art in aspects such as minimal angle, vertex valence and triangle quality.
We discuss why remeshing benefits from blue noise properties, maximal sampling, bounds on vertex valence, and geometric bounds (e.g., angle bounds). These properties are important to simulation applications. For example, Schechter and Bridson~\shortcite{Schechter2012} demonstrated that the simulation result based on Poisson-disk sampling performs much better than that based on the regular grid sampling. We propose a remeshing framework that jointly optimizes these criteria and evaluate it in a comparison to several recent remeshing algorithms.
The main contributions of these three parts are:
\begin{itemize}
\item A simple and elegant theoretical analysis of the gap geometry that improves the work of~\cite{Ebeida2011} for 2D, and is the first analysis for \ndim{d} and manifolds.
\item The design of efficient algorithms and data structures for all the gap processing operations identified above.
\item A surface remeshing (as well as 2D meshing) algorithm that compares favorably to the state of the art in aspects such as minimal angle, vertex valence and triangle quality.
\end{itemize}

\begin{figure}
\centerline{
\includegraphics[width=1.0\linewidth]{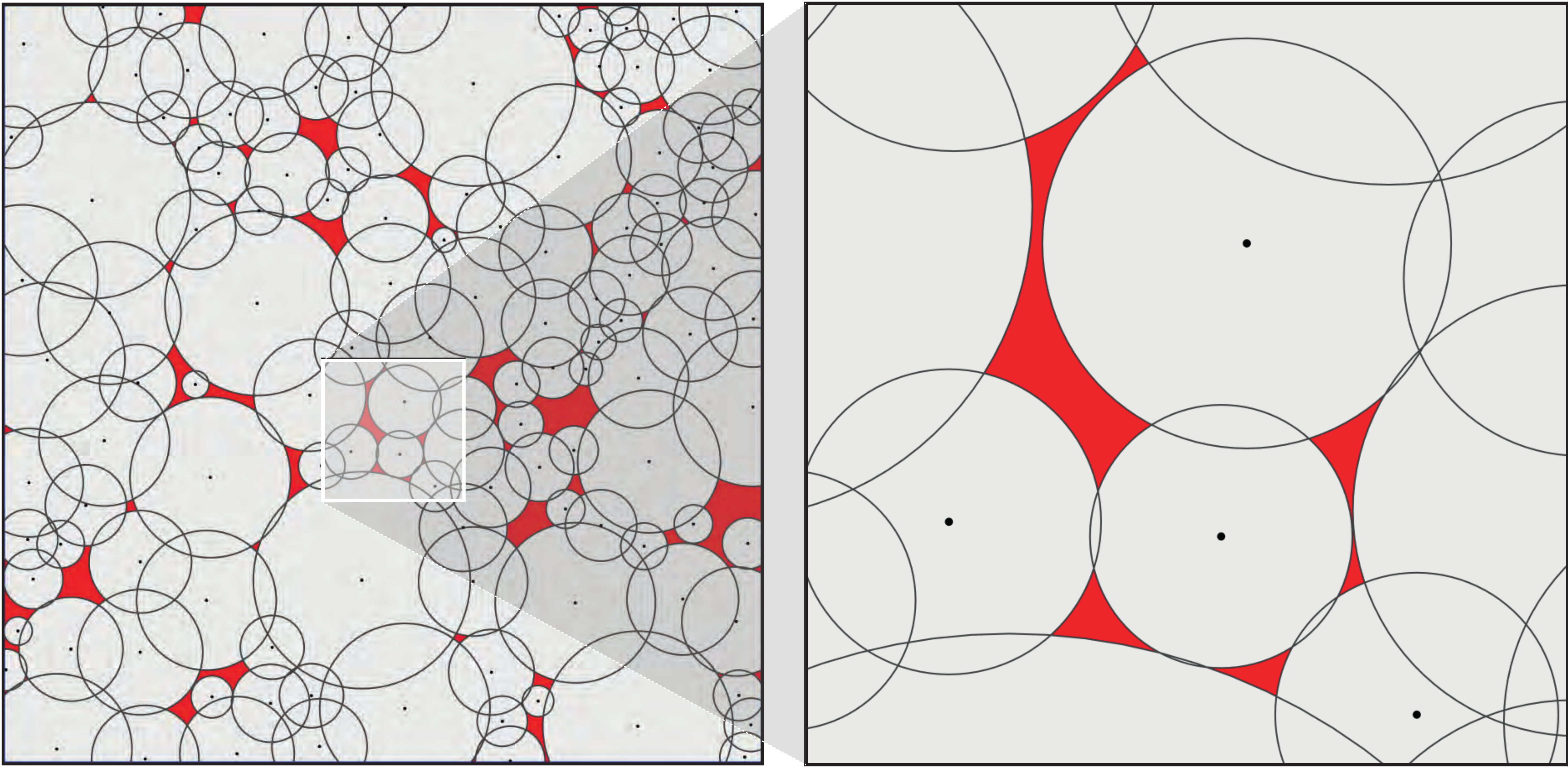}
}
\caption{Left: Illustration of gaps of a disk set; the uncovered regions are shown in red. Right: zoom in view of the uncovered regions.}
\label{fig:tinygap}
\end{figure}

\subsection{Related work}

We briefly review techniques for generating Poisson-disk point sets, but refer the reader to~\cite{Lagae2008} for a more comprehensive survey. Our review focuses on the treatment of gaps in these algorithms.

There are a large number of sampling algorithms that produce point sets with different attributes~\cite{Lloyd1982,Dippe1985,Cook1986,Mitchell1987,Mitchell1991,Mccool1992,Turk1993} that are either not concerned with the geometry of gaps or do not use an acceleration data structure. Our work can improve upon some of these methods, e.g., ~\cite{Mccool1992}.

\paragraph*{Poisson-disk sampling}
Most recent algorithms for efficient Poisson-disk sampling maintain a data structure of gap primitives. The simplest primitive is a square, because it is easy to sample~\cite{White2007,Bridson2007,Gamito2009,Ebeida2012} and it can be efficiently subdivided. These algorithms are extremely efficient when trying to fill a large mainly uncovered region, but even when filling small gaps, they are efficient enough so that the overall running time is much faster than the algorithm built on an exact representation of gaps~\cite{Dunbar2006}. Ebeida et al.~\shortcite{Ebeida2011} propose a hybrid approach that first uses squares and later convex polygons bounding the intersections of a square and multiple circles as gap primitives. A two-step Poisson-disk sampling framework is proposed that first performs dart-throwing in the grid and then against the extracted gaps. The maximal property is achieved and the sampling framework is shown to be more efficient then previous approaches. Jones~\shortcite{Jones2006} uses a Voronoi diagram to extract gap primitives. The algorithm repeatedly generates disks and inserts them in a global Voronoi diagram one by one. Each vertex in the Voronoi diagram maintains a value that indicates the area of empty region of the corresponding Voronoi cell. A new sampling is generated by first selecting a Voronoi cell based on the empty area. The Voronoi diagram and the value of each vertex is updated each time a new sample is generated. Jones' algorithm uses the fact that a maximal sampling can be obtained if and only if the Voronoi cell of each vertex is fully covered by the disk centered at the vertex. However, Jones~\shortcite{Jones2006} as well as the other approaches discussed above handle only 2D uniform sampling. In contrast, we use a power diagram and the dual regular triangulation for our analysis, which results in a more general and much simpler solution.

\paragraph*{Adaptive sampling} Fattal presents an adaptive sampling algorithm based on kernel density estimation~\cite{Fattal2011}. Kalantari et al.~\shortcite{Kalantari2011} propose to use joint distributions by breaking down the 2D probability density function (pdf) into a 1D conditional pdf (cpdf) and a 1D pdf for maximal adaptive sampling. However, the proposed algorithm is dependent on a threshold for discretizing the 1D pdf, which cannot achieve maximal sampling. In concurrent work, Mitchell et al.~\shortcite{Mitchell2012} study 2D Poisson-disk sampling with various radii, and de Goes et al. formulate the blue noise sampling problem using optimal transport~\cite{deGoes2012}. However, the maximality is not discussed in these approaches.

\paragraph*{Sampling on surfaces}
Fu and Zhou~\shortcite{Fu2008} generalize scalloped sector based sampling~\cite{Dunbar2006} to 3D mesh surfaces, and present an isotropic remeshing algorithm by extracting a mesh from the samples. Lloyd iterations are used to further smooth the resulting mesh. Cline et al.~\shortcite{Cline2009} propose dart throwing algorithms on surfaces based on a hierarchical triangulation. Bowers et al.~\shortcite{Bowers2010} extend the parallel sampling method ~\cite{Wei2008} to mesh surfaces and introduce a spectral analysis for uniform surface sampling algorithms as well. Wei and Wang ~\shortcite{Wei2011} present a framework for spectral analysis of nonuniform blue noise sampling for both 2D domains and surfaces. Again, none of these methods satisfy the maximal sampling property. Xu et al.~\shortcite{Xu2012} extend the concept of \emph{Capacity-Constrained Voronoi Tessellation} (CCVT)~\cite{Balzer2009} by introducing capacity-constrained Delaunay triangulations for blue noise sampling on surfaces. Chen et al.~\shortcite{Chen2012} combine the CCVT~\cite{Balzer2009} and the \emph{Centroidal Voronoi Tessellation} (CVT) framework~\cite{Yan2009} for blue noise sampling on surfaces. However, the  approaches of Chen et al.~\shortcite{Chen2012} and Xu et al.~\shortcite{Xu2012} are also based on the Lloyd iterations. Corsini et al.~\shortcite{Corsini2012} present an algorithm for surface blue noise sampling based on a space subdivision combined with a pre-generation of the samples. Although many surface blue noise sampling algorithms have been proposed recently, few of them use blue noise sampling for geometric applications, such as remeshing.

\paragraph*{Other approaches}
There are several extensions to Poisson-disk sampling that we did not consider in our current framework, but that might be interesting avenues for future work, e.g., parallel sampling~\cite{Wei2008,Gamito2009} and multi-class sampling~\cite{Wei2010}.
Schl\"omer et al.~\shortcite{Schlomer2011b} proposed the use of the farthest Voronoi diagram for blue noise sampling. Some approaches pre-compute tile sets and then quickly arrange them in realtime~\cite{Ostromoukhov2004,Kopf2006}. Our work might be beneficial for the tile pre-computation phase.

\section{Theoretical Gap Analysis}
\label{sec:overview}

We study a set of disks $\cD=\{(\mp_i, r_i)\}_{i=1}^n$ in a compact \ndim{d} domain $\Omega$, where $\mp_i$ and $r_i$ are the center and the radius of the $i^{th}$ disk, respectively. Here, we use the word \emph{disk} to represent a general disk, i.e., a sphere in 3D and a hyper-sphere in \ndim{d} ($d>3$). We assume that the center $\mp_i$ of any disk cannot be covered by the other disks, i.e., $\forall{i,j}\,(i\neq\,j),\|\mp_i-\mp_j\|\geq \max(r_i,r_j)$. If we draw a disk at each center $\mp_i$ with radius $r_i$, the domain $\Omega$ will be fully or partially covered. We are interested in the properties of the uncovered region, which is defined as $\Omega-\cD$. As shown in Figure~\ref{fig:tinygap}, if the domain $\Omega$ is partially covered by a set of disks, the uncovered region is split into a set of isolated regions. We use the term \emph{gap} to refer to a connected component of the uncovered region.

At first glance, a gap can be arbitrarily complex or can be arbitrarily small so that it is very difficult to sample gaps directly. Therefore, like most previous work, we shall decompose these complex gaps to a set of gap primitives $GP_i$ such that 1) the union of these gap primitives covers the gaps $\Omega-\cD\subseteq\cup{GP_i}$, 2) the gap primitives are non-intersecting $\forall{i, j}\,(i\neq{j}), GP_i\cap{GP_j}=\emptyset$, and 3) it is easy to sample a gap primitive using rejection sampling.

\subsection{Power diagram and regular triangulation}

\begin{figure}[t]
\centerline{
\hfill
\includegraphics[width=0.453\linewidth]{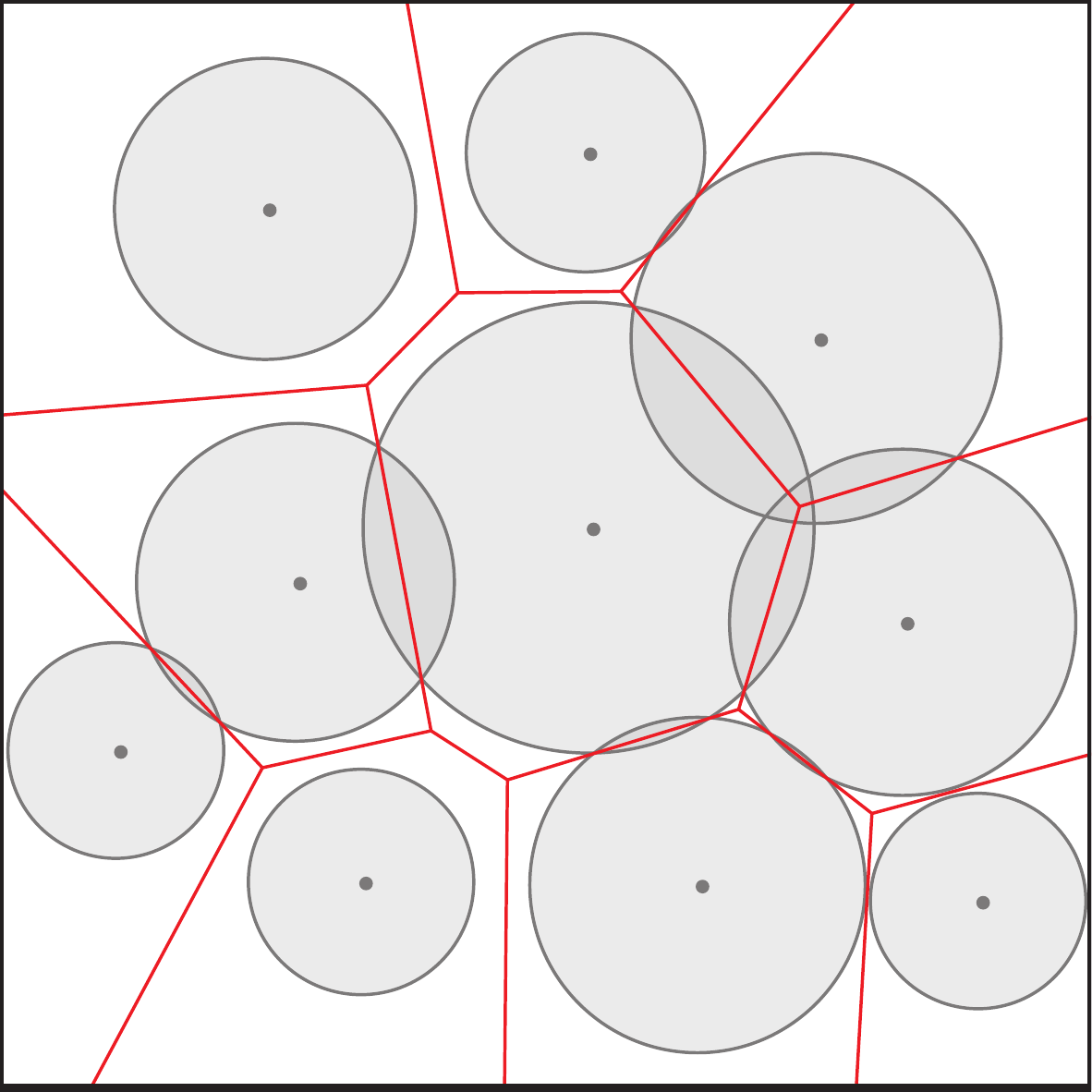}
\hfill
\includegraphics[width=0.453\linewidth]{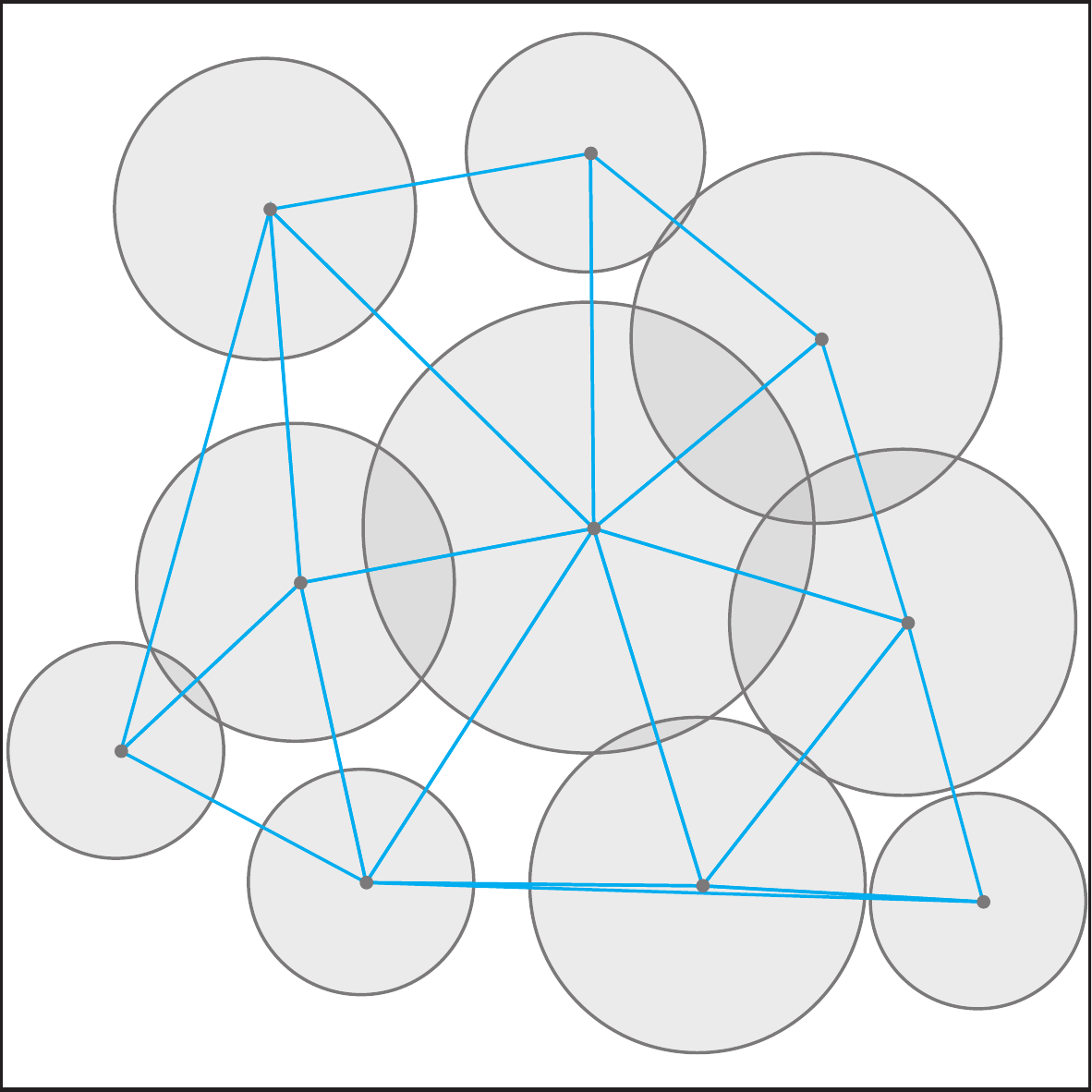}
\hfill
}
\caption{The power diagram (left) and the regular triangulation (right) of a set of disks.}
\label{fig:pdrt}
\end{figure}

We analyze the geometric properties of gaps based on the concept of the power diagram~\cite{Aurenhammer1987}, and its dual, called regular triangulation.
The power diagram is a tessellation of the Euclidean space into a set of convex polygons defined by a set of disks. Figure~\ref{fig:pdrt} shows an example of the power diagram of a set of disks. The power diagram and the regular triangulation are equivalent to the Voronoi diagram and the Delaunay triangulation when the radius of all the points are the same.

The disk set $\cD$ is represented by a set of weighted points $\wps = \{(\mp_i, w_i)\}_{i=1}^n$, where $w_i = r_i^2$. The power of two weighted points is defined as $$\Pi(\mp_i,w_i,\mp_j,w_j)=\|\mp_i-\mp_j\|^2-w_i-w_j.$$
Then, the power diagram $\pd$ of $\wps$ is a set of non-overlapping power cells $\{\Omega_i\}_{i=1}^n$, such that
$$\Omega_i=\{\mx\in\Omega \mid \Pi(\mp_i,w_i,\mx,0)\leq\Pi(\mp_j,w_j,\mx,0), \forall j\neq\,i\}.$$
A vertex of the power diagram is called a \emph{power vertex} and an edge is called a \emph{power edge}. A \ndim{d} power diagram can be interpreted as the intersection of a \ndim{(d+1)} Voronoi diagram and a \ndim{d} hyperplane~\cite{Ash1986}.

\ifx\highdim\nodefined
The regular triangulation $\rt$ of $\wps$ is the dual of the power diagram, which consists of a set of triangles, i.e., $\rt(\wps)=\{t_j\}$. Each triangle $t\in\rt$ is formed by three weighted points, such that there exists a point $\mc_t\in\Omega$ that has equal powers \wrt the three vertices of $t$, and this power is less than the power of $\mc_t$ \wrt any other weighted points in the triangulation. Each $\mc_t$ is called the \emph{power center} of $t$, which is a power vertex of $\pd$. Each edge of the regular triangulation is dual to a power edge. We denote the power of a triangle $t$ by $\Pi(t) = \Pi(\mp, w, \mc_t, 0)$, where $\mp$ refers to one of the triangle's vertices, and $w$ is the corresponding weight of $\mp$. The concept of the power diagram and the regular triangulation are related to that of the Voronoi diagram and the Delaunay triangulation.
\else
The regular triangulation of $\wps$ is a \ndim{d} simplicial complex $\cT$ and it is the dual of the power diagram. A \ndim{d} simplicial complex is the union of a set of $k$-simplices ($0\leq\,k\leq\,d$). Each $k$-simplex is the convex hull of $k+1$ linear independent points that are dual to a \ndim{(d-k)} convex polyhedron of the power diagram. The simplicial complex $\cT$ satisfies: i) any face of a simplex is also in $\cT$, and ii) the intersection of any two simplices is either empty, or is a face of both simplices. We denote the set of the $d$-simplices (i.e., triangles in 2D and tetrahedra in 3D) as $\rt=\{t_j\}$.
For each $t\in\rt$, there exists a point $\mc_t\in\Omega$  that has equal powers \wrt the $d+1$ vertices of $t$, and this power is less than the power of $\mc_t$ \wrt any other weighted points in the triangulation. Each $\mc_t$ is called the \emph{power center} of $t$, which is a power vertex of $\pd$. We denote the power of a $d$-simplex $t$ by $\Pi(t) = \Pi(\mp, w, \mc_t, 0)$, where $\mp$ refers to one of the vertices of $t$, and $w$ is the corresponding weight of $\mp$.
\fi

\begin{figure}[t]
\centerline{
\includegraphics[width=0.98\linewidth]{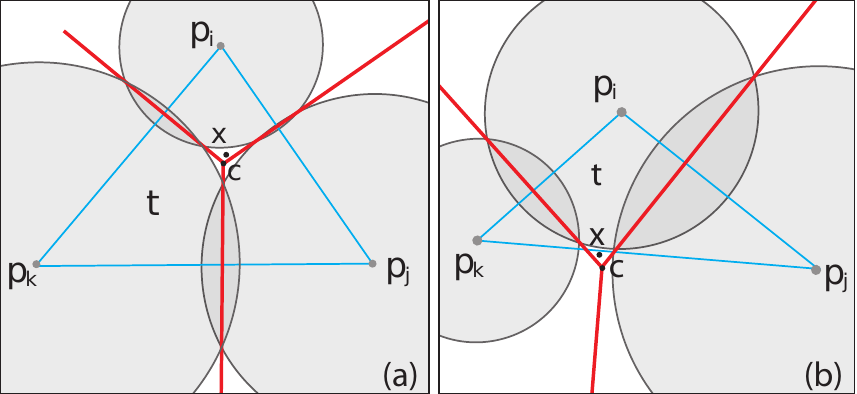}
}
\caption{Existence of gaps in the 2D plane. The power center $\mc$ of a triangle $\triangle{\mp_i\,\mp_j\,\mp_k}$ can be used to test for the existence of gaps.}
\label{fig:proof}
\end{figure}

\ifx\highdim\nodefined
\subsection {Existence of gaps} \label{sec:gapexist}
In the following, we give the theorems that state the condition of the existence of gaps in a 2D plane, and on 3D surfaces.

\begin{theorem}
A gap exists iff $\exists\,t\in{\rt},\Pi(t)>0$.
\label{thm:gap}
\end{theorem}

\else
\subsection {Existence of gaps} \label{sec:gapexist}
In the following, we give the theorems that state the condition of the existence of gaps in \ndim{d} spaces and on 3D surfaces, respectively.

\begin{theorem}
A gap exists iff $\exists\,t\in{\rt},\Pi(t)>0$.
\label{thm:gap}
\end{theorem}

Here, we give the proof only in 2D (see Figure~\ref{fig:proof}), as it is more intuitive and the proof for higher dimensions is similar.
\fi

\begin{proof}

$\Rightarrow$ Suppose that there is a gap. For any point $\mx$ in this gap, we define the function $F(\mx)$ as the minimal power between the point $\mx$ and the weighted points $\wps$. Suppose that $\mx$ falls in the power cell $\Omega_i$ of the sample $(\mp_i,w_i)$. Clearly, $F(\mx)=\Pi(\mx, \mp_i)$ by definition of the power diagram. $F(\mx)$ attains its maxima when $\mx$ coincides with a power vertex $\mc$, which is the power center of a triangle $t\in\rt$. When $\mx$ is in the gap, by definition $F(\mx)>0$ since $\Pi(\mx,\mp_i)>0$. Note that $\Pi(t)=\Pi(\mc, \mp_i)=\max(F(\mx))$. Hence, $\Pi(t)>F(\mx)>0$.

$\Leftarrow$ If $\exists{t}$, $\Pi(t)>0$, then the power center of the triangle $t$ is not covered. Therefore, there exists a gap.
\end{proof}

\paragraph*{Gaps on 3D surfaces}
If the input domain $\Omega$ is a 3D surface, Thm.~\ref{thm:gap} does not apply since the domain is not flat. In this case, we generalize the approach of Jones~\shortcite{Jones2006} to 3D surfaces using the concept of the \emph{Restricted Power Diagram} ($\rpd$).

The restricted power diagram is a generalization of the \emph{Restricted Voronoi Diagram} (RVD)~\cite{Edelsbrunner1997} on surfaces. The $\rpd$ is defined as the intersection of the 3D power diagram and the surface, i.e., $\rpd(\wps)=\pd(\wps)\bigcap\Omega=\{\Omega_i\bigcap\Omega\}_{i=1}^n$.

\begin{theorem}
A gap in the disk set $\cD$ (sphere set in 3D) on a 3D surface exists iff there exists a sample $\mp_i$ whose corresponding restricted power cell is not fully covered by the sphere $(\mp_i,r_i)$ centered at $\mp_i$.
\label{thm:gapsurf}
\end{theorem}
Thm.~\ref{thm:gapsurf} describes the global condition of gap existence on surfaces. The proof is straightforward and we omit it here. In Section~\ref{sec:remesh}, we show how to locally detect gaps on mesh surfaces in the context of $\epsilon$-sampling and \emph{Restricted Regular Triangulations} ($\rrt$).

\section{2D Gap Primitive Processing} \label{sec:gap2d}

In the following, we describe the most important operations related to gap processing in the 2D plane. In the following discussion, a triangle $t$ is called a \emph{gap triangle} if $\Pi(t)>0$, or in other words, the power center of the triangle is not covered by the three disks centered on the vertices of the triangle. We provide a short overview of each of these operations here and in the subsections we describe each operation in more detail:
\begin{itemize}
\item Gap detection: gap detection refers to analyzing an existing point set to determine if gaps exist and where they are. If no gap exists, we can determine that a point set is maximal. In our solution, this operation outputs a list of gap triangles (see Section~\ref{sec:detect}).
\item Gap clustering: this operation groups all gap triangles belonging to the same independent gap set (Section~\ref{sec:cluster}) together. In our proposed solution, this clustering is performed on the level of gap triangles, before the actual gaps are extracted. This step is optional but it enables a simple parallelization of gap sampling.
\item Gap primitive extraction: given a set of input points, this operation computes a set of gap primitives covering all gaps (Section~\ref{sec:extract}).
\item Gap primitive updates: these operations update the gap primitives after points are inserted, deleted, moved, or when one or multiple disk radii are changed (Section~\ref{sec:update}).
\end{itemize}
The state of the framework is the weighted point set $\wps$ and the corresponding regular triangulation $\rt$. At the beginning, $\rt$ has to be initialized once.

\subsection{Gap detection} \label{sec:detect}

We can optionally collect all the gap triangles in an array for later use or simply return a Boolean to indicate the existence of gaps.
Note that a gap triangle does not necessarily mean that the triangle itself contains some uncovered region, since its power center can be outside the triangle. See Figure~\ref{fig:vis}(b) for an example where the triangle $t_0$ is fully covered but $\Pi(t_0)>0$. The incident gap region is outside of the triangle $t_0$. Also, since a gap might contain several triangle power centers, one gap can have several corresponding gap triangles.
We traverse all the triangles in $\rt$, compute the power center $c_i$ and power $\Pi(t_i)$ of each triangle $t_i$. If $\Pi(t_i)>0$, triangle $t_i$ is marked as a \emph{gap triangle} according to Thm.~\ref{thm:gap}.

\subsection{Gap clustering}
\label{sec:cluster}

We say that two gaps are dependent, if a disk inserted in one gap might intersect the other gap. We cluster all dependent gaps into \emph{independent gap sets} (IGS), which can be used later for parallel sampling. With slight abuse of notation, we sometimes also use IGS to refer to the corresponding collection of gap triangles or gap primitives.

\begin{figure}[htbp]
\centerline{
\includegraphics[width=0.45\linewidth]{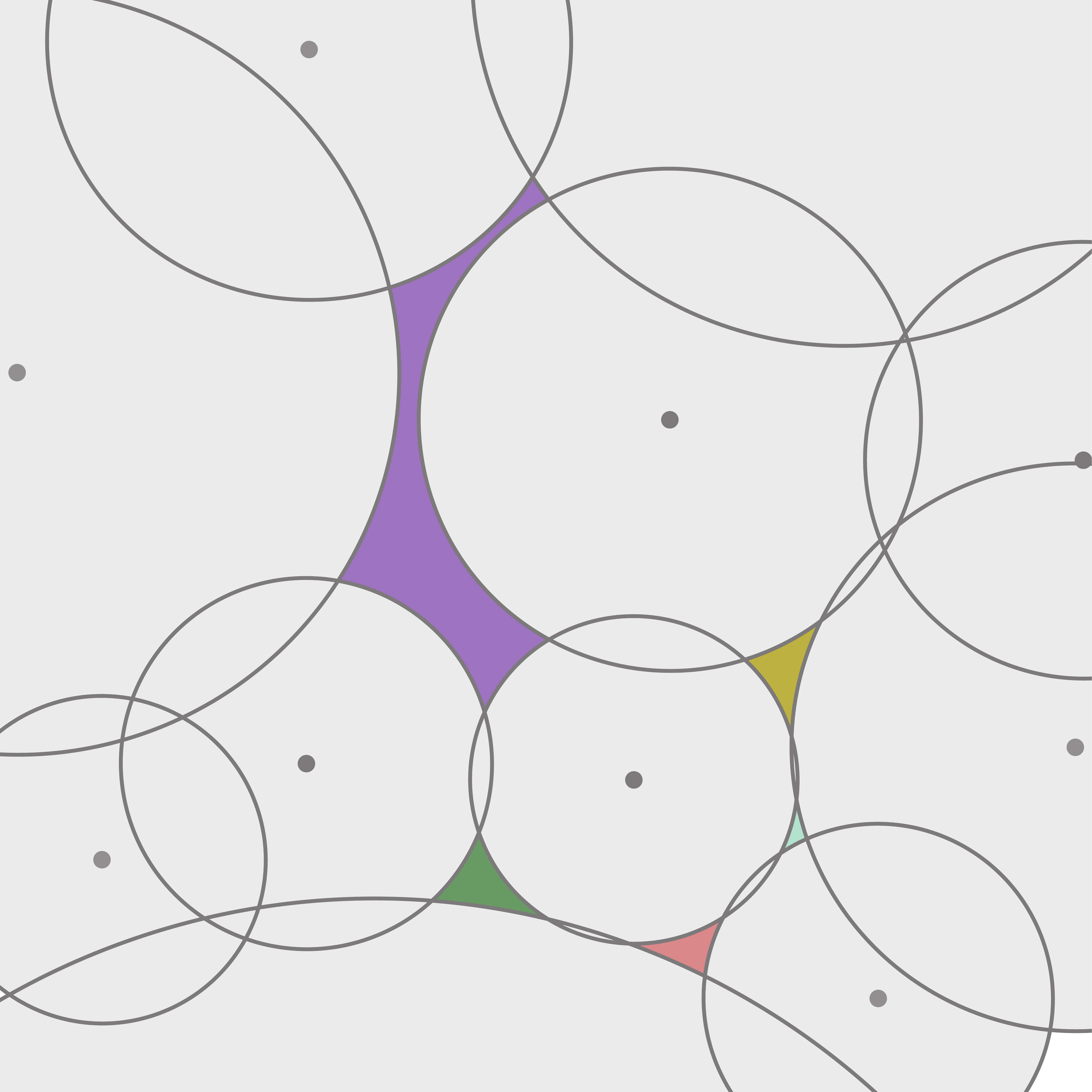}
\hfill
\includegraphics[width=0.45\linewidth]{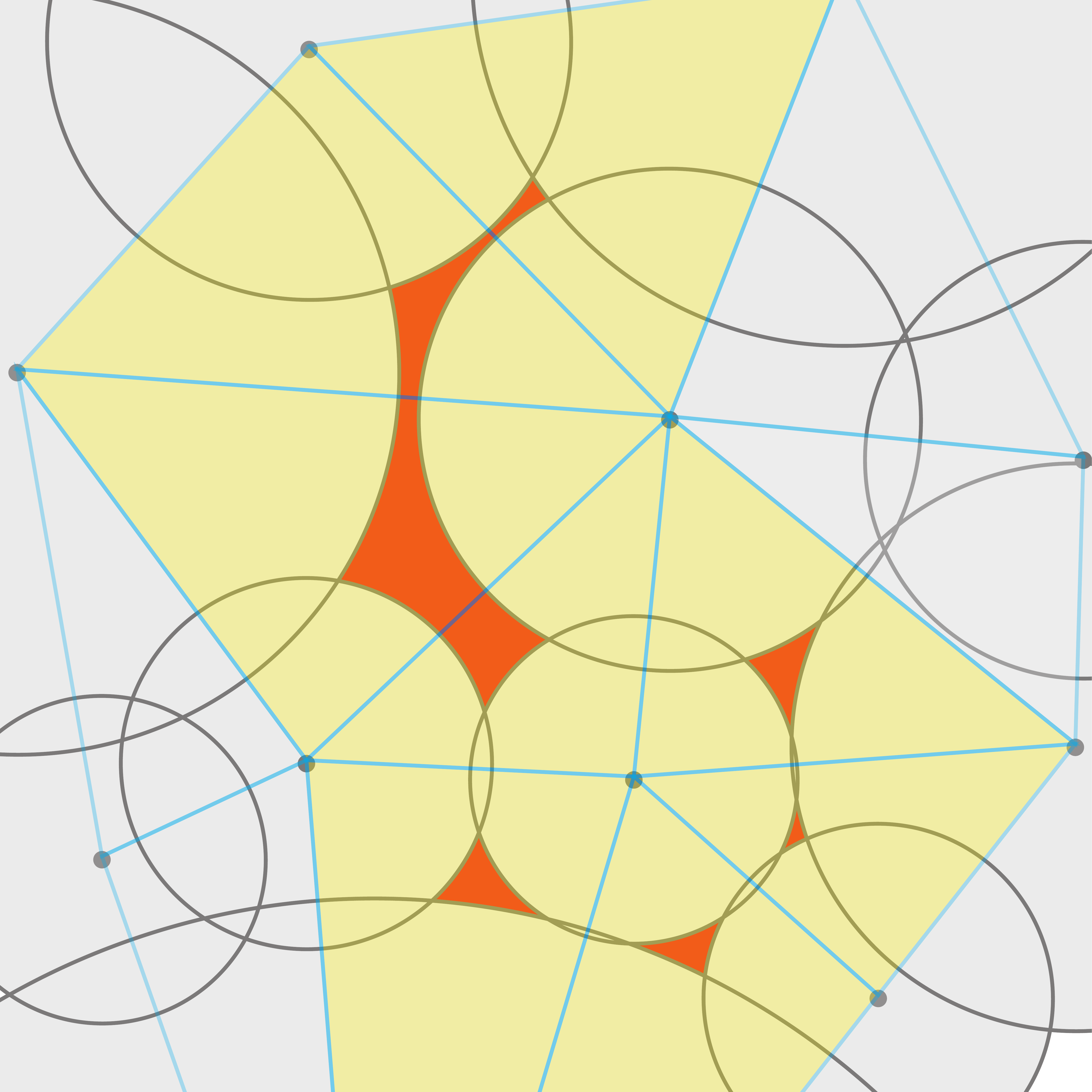}
}
\caption{Gap clustering. There are five gaps (left) that clustered together forming an independent gap set (right) by the clustering algorithm.}
\label{fig:cluster}
\end{figure}

It is clear that connected gaps are dependent. However, even disconnected gaps can be dependent if the sampling radius inside the gaps is larger than the distance between the gaps. This is only a sufficient condition, because additionally a disk can only be placed if it does not cover another existing disk center. Since it is difficult to exactly compute the dependence of gaps, we propose a conservative approach to compute IGS clustering: all gap triangles in the one-ring neighborhood of each other are clustered in the same IGS (Figure~\ref{fig:cluster}). While we checked that the clustering is conservative in praxis, we do not offer a formal proof in this paper.

In our implementation, the algorithm is performed for gap triangles, but there is a simple one-to-one correspondence to gap primitives (see Section~\ref{sec:extract}). All the gap triangles are marked as unvisited at the beginning. We then traverse all the gap triangles. Each time when we encounter an unvisited triangle, we extract the IGS using a simple region growing algorithm, i.e., if two neighboring triangles are gap triangles, then they belong to the same group of an IGS.
The grouped triangles are marked as visited and the traversal continues. This step ends when all the gap triangles are marked as visited.
There are three cases of the relationship between two neighboring gap triangles, which are called \emph{connectivity rules.} To cluster IGSs, all three rules apply, but if only connected components are desired, the last rule does not apply.

\begin{figure}[t]
\centerline{
\includegraphics[width=0.5\linewidth]{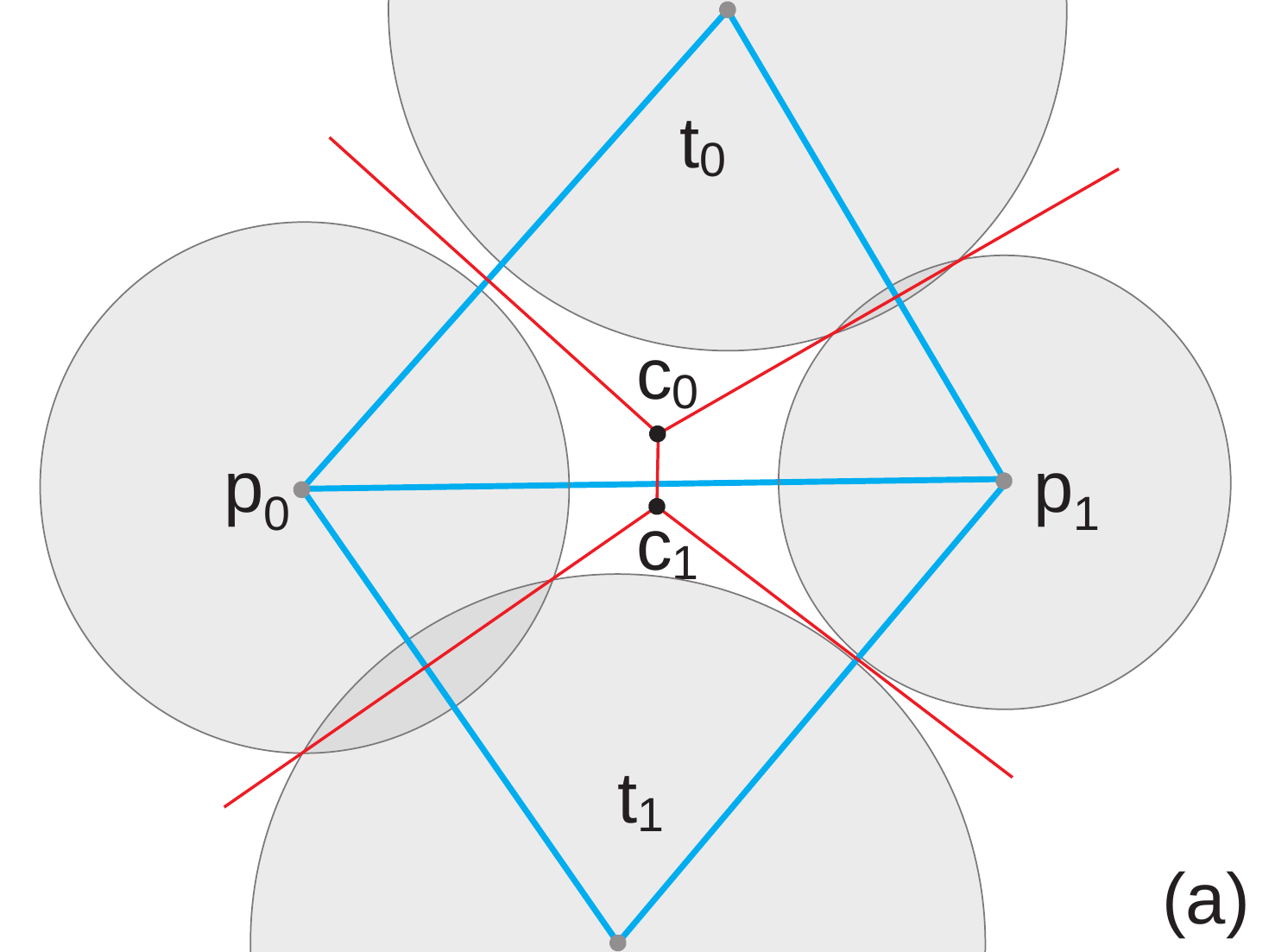}
\includegraphics[width=0.5\linewidth]{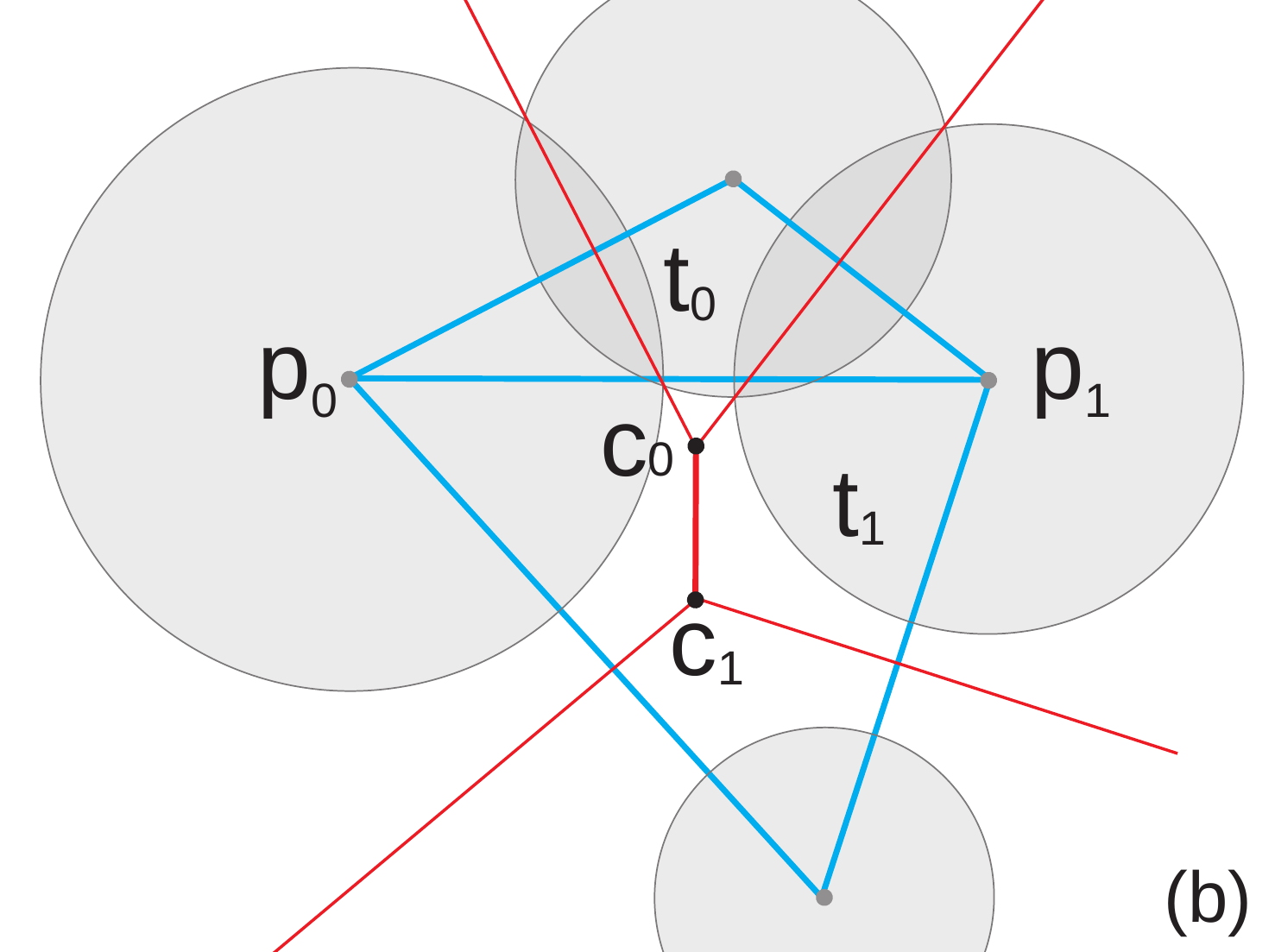}
}
\centerline{
\includegraphics[width=0.5\linewidth]{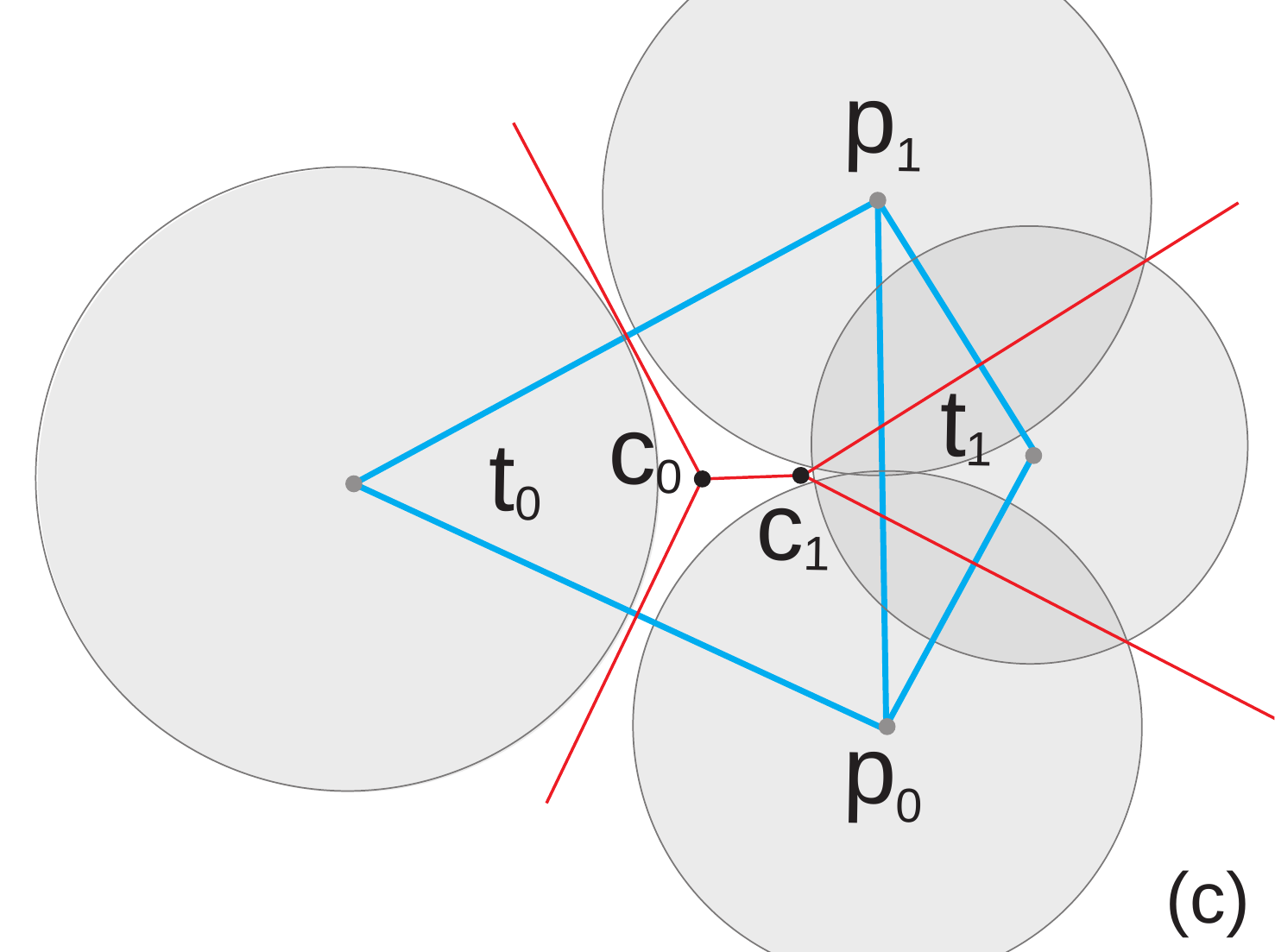}
\includegraphics[width=0.5\linewidth]{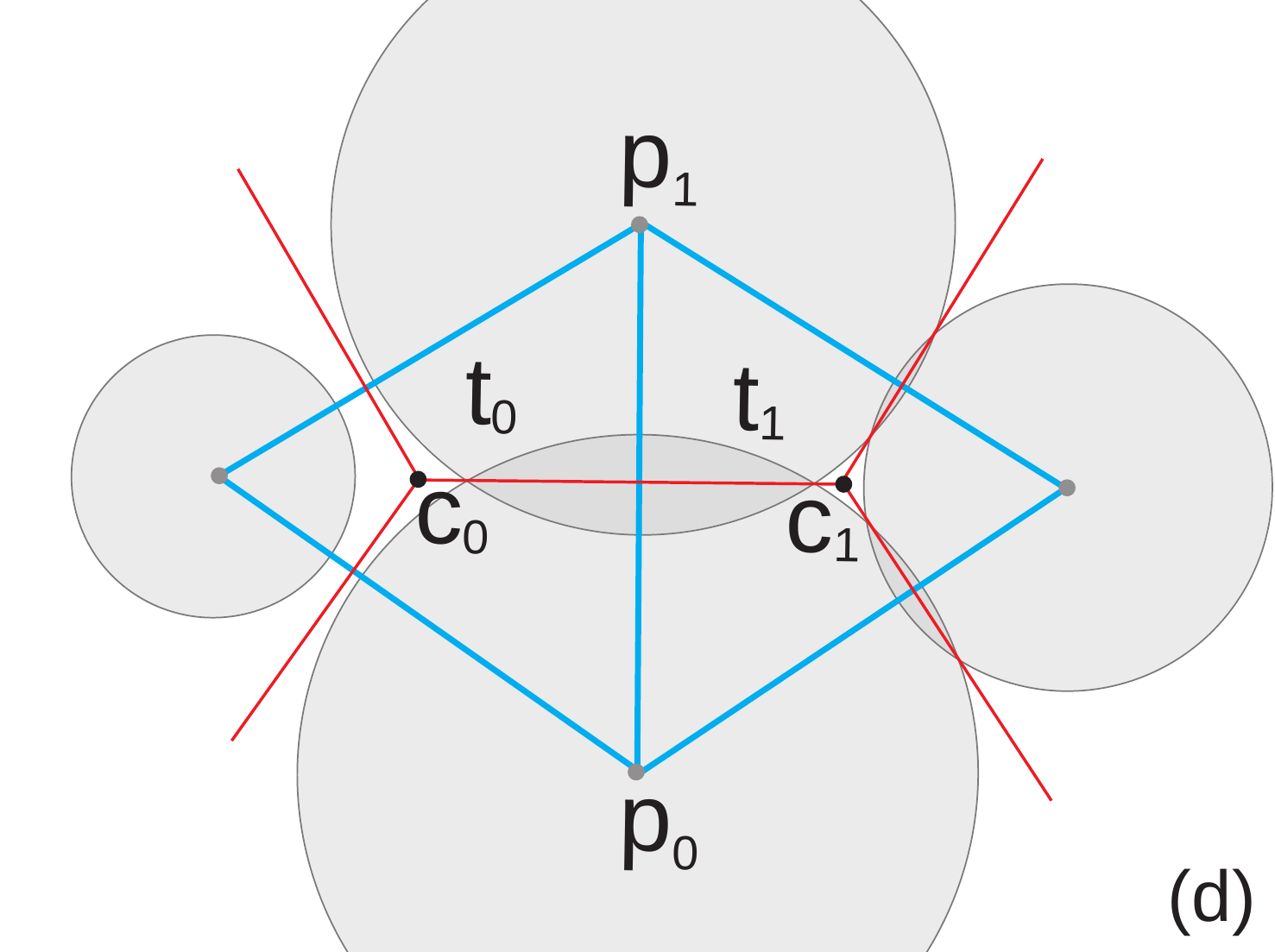}
}
\caption{Illustration of the connectivity between two neighboring gap triangles $t_0, t_1$ (blue triangles) where $c_0, c_1$ are the corresponding power centers, respectively. $\mp_0\,\mp_1$ are on the common edge shared by $t_0, t_1$. (a)~$|\mp_0\,\mp_1|>r_0+r_1$, and  $c_0, c_1$ are on the different sides of $\mp_0\,\mp_1$. (b)~$|\mp_0\,\mp_1|>r_0+r_1$, and $c_0, c_1$ are on the same side of $\mp_0\,\mp_1$. (c)~$|\mp_0\,\mp_1|<r_0+r_1$, and $c_0, c_1$ are on the same side of $\mp_0\,\mp_1$. (d)~$|\mp_0\,\mp_1|<r_0+r_1$, but $c_0, c_1$ are on the different sides of $\mp_0\,\mp_1$. }
\label{fig:vis}
\end{figure}

\paragraph*{Connectivity rules} Given a pair of neighboring gap triangles $t_0, t_1$, and their power centers $\mc_0, \mc_1$, respectively (Figure~\ref{fig:vis}), we use the following connectivity rules:
\begin{itemize}
\item The length of the common edge shared by $t_0, t_1$ is larger than $r_0+r_1$, where $r_0, r_1$ are the corresponding sampling radii at the vertices of the common edge $\mp_0,\mp_1$ (Figure~\ref{fig:vis}(a) and~\ref{fig:vis}(b)).
\item The length of the common edge shared by $t_0, t_1$ is smaller than $r_0+r_1$, but $\mc_0, \mc_1$ are on the same side of the common edge (Figure~\ref{fig:vis}(c)).
\item The length of the common edge shared by $t_0, t_1$ is smaller than $r_0+r_1$, and $\mc_0, \mc_1$ are on different sides of the common edge (Figure~\ref{fig:vis}(d)).
    In this case, although two gaps are disconnected, putting a new disk in one empty region may affect the geometry on the other side. The two triangles are therefore classified as belonging to the same IGS.
\end{itemize}

\begin{figure}[t]
\centerline{
\subfigure[triangle]{\includegraphics[width=0.33\linewidth]{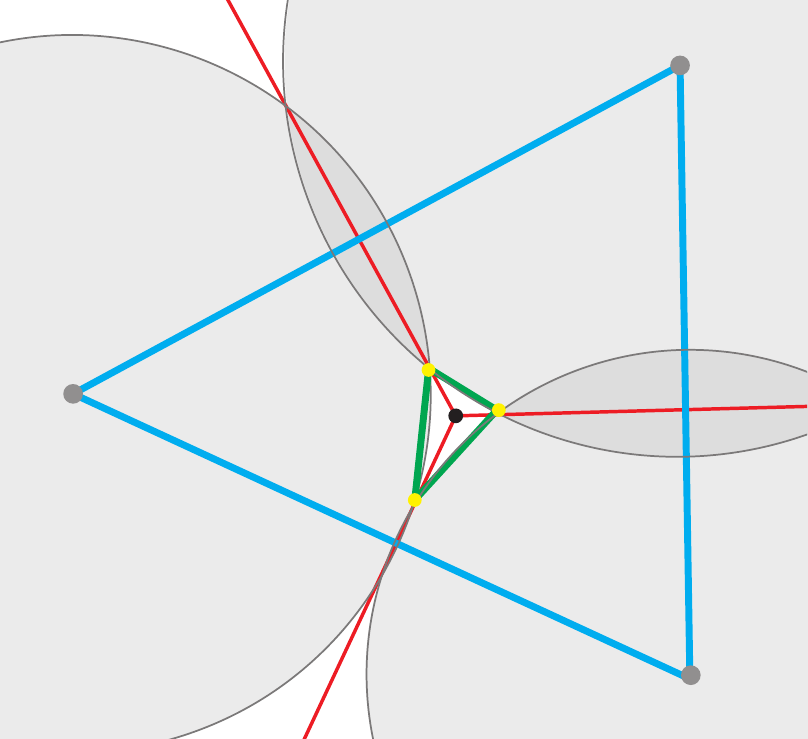}}
\subfigure[quad]{\includegraphics[width=0.33\linewidth]{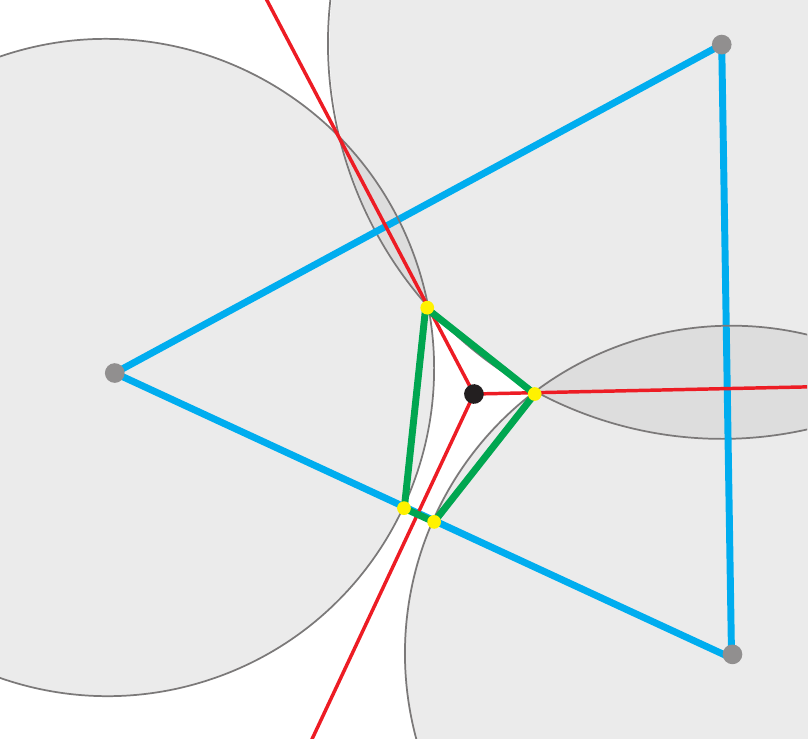}}
\subfigure[pentagon]{\includegraphics[width=0.33\linewidth]{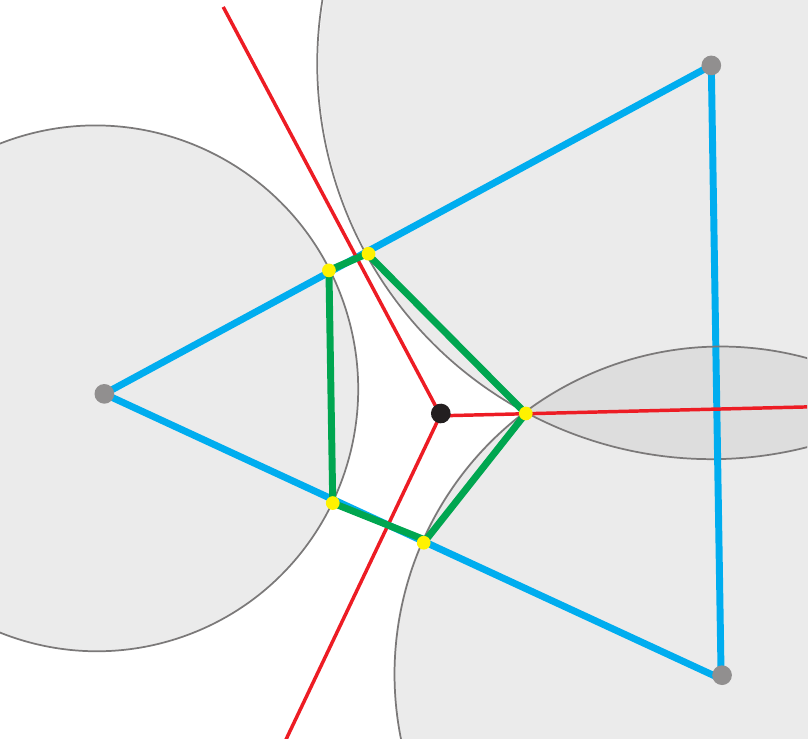}}
}
\centerline{
\subfigure[hexagon]{\includegraphics[width=0.33\linewidth]{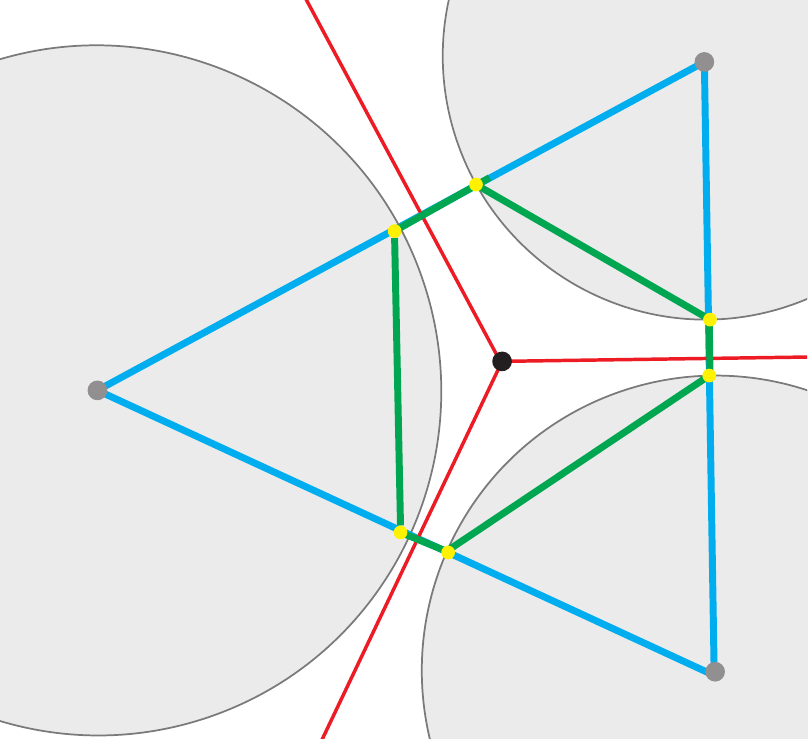}}
\hfill
\subfigure[special case]{\includegraphics[width=0.66\linewidth]{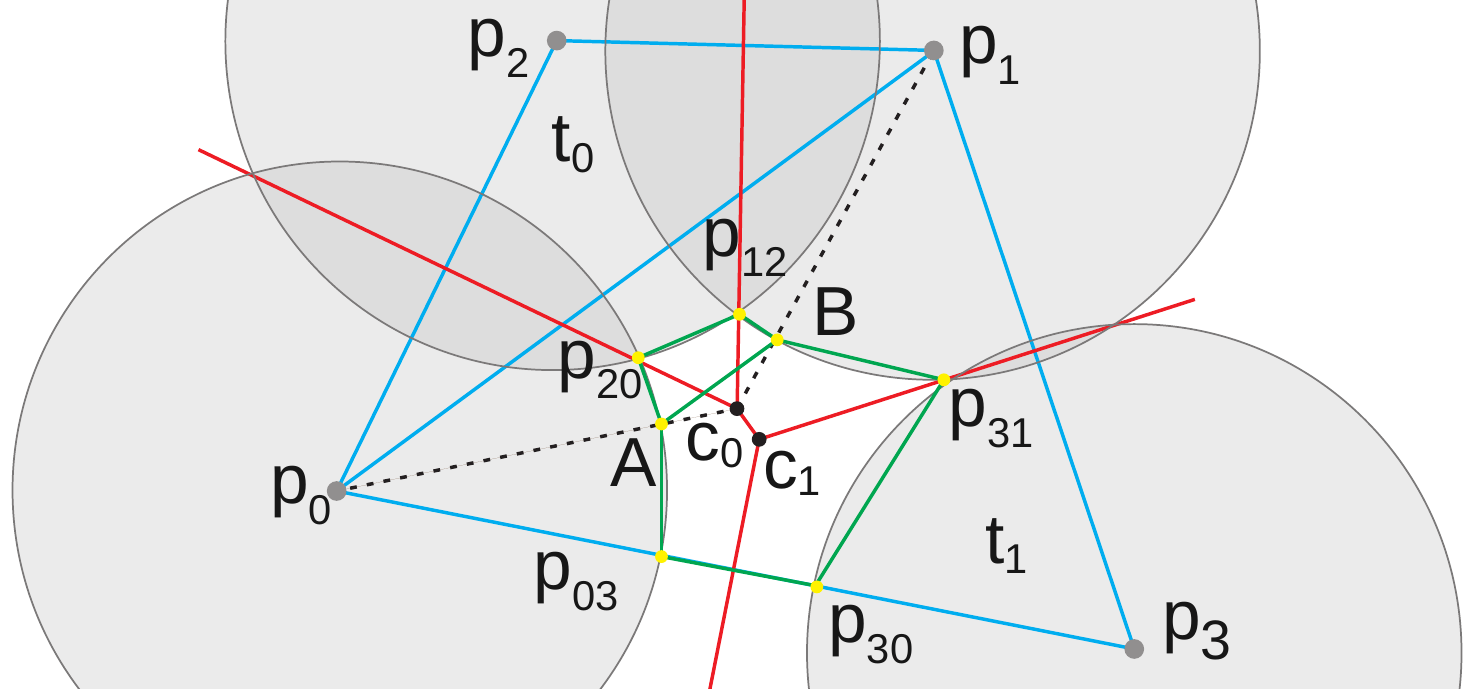}}
}
\caption{(a)-(d) Simple gap primitives contained inside the triangle,
and (e) a special case when the power center is outside the triangle. }
\label{fig:pri}
\end{figure}

\subsection{Gap primitive extraction}
\label{sec:extract}

The algorithm extracts a \emph{gap primitive} for each triangle of the gap separately. Each gap primitive is a simple convex polygon with up to six edges. The \emph{connectivity rules} are used to assist the extraction. Given a triangle $t_0$ belonging to a gap, we simply traverse the three neighbors of $t_0$ and extract the vertices of the gap primitive based on the classification of the types of neighboring triangles. We assume that $t_1$ is one of $t_0$'s neighbors, and $\mc_0, \mc_1$ are power centers of $t_0$ and $t_1$, respectively. $\mp_0\,\mp_1$ is the common edge shared by $t_0$ and $t_1$, and the direction of $\mp_0\,\mp_1$ is \emph{ccw} (counter-clockwise) in triangle $t_0$.
Then, there can be the following two cases:

First, if the triangle $t_0$ contains only the power center $\mc_0$ of itself (see Figure~\ref{fig:pri}(a)-(d)), then we can extract the gap primitive directly by traversing its three (ccw) edges using the following rules:
\begin{itemize}
\item If $|\mp_0\,\mp_1|\leq{r_0+r_1}$, we compute the intersection points of the disks centered at $\mp_0$ and $\mp_1$ and the intersection point on the left side of $\mp_0\,\mp_1$ is added to the gap primitive of $t_0$.
\item If $|\mp_0\,\mp_1|>r_0+r_1$, we compute the intersection points of the disks and edge $\mp_0\,\mp_1$ and the resulting points $\mp_{01}$ and $\mp_{10}$ are added to the gap primitive of $t_0$.
\end{itemize}

Second, if the triangle $t_0$'s power center is outside $t_0$, e.g., if it lies in its neighboring triangle $t_1$, then all these triangles belong to the same connected gap component according to the connectivity rules. Note that in complex cases, $t_0$'s power center can also lie in its neighbor's neighboring triangle. This case is equivalent to that of a triangle containing more than one power center. To extract gap primitives is equivalent to decomposing the gap w.r.t. each gap triangle. We handle this configuration as follows:
\begin{itemize}
\item For the edges we currently traversed, if the power centers of two incident triangles lie on two different sides, we compute the intersection points as before.
\item For the edges with two power centers of two incident triangles lying on the same side, e.g., edge $\mp_0\,\mp_1$ shown in Figure~\ref{fig:pri}(e), we simply connect $t_0$'s power center $\mc_0$ and $\mp_0$, $\mp_1$, which results in two intersection points A and B. The segment AB is then added to the primitive polygon of $t_0$. Similarly, if the neighboring triangle's power center lies on the same side of of the common edge of the current triangle, e.g., $t_1$ in Figure~\ref{fig:pri}(e), we also add the segment BA to $t_1$'s gap primitive.
\end{itemize}

The above extraction process decomposes a connected gap component into non-overlapping simple polygons. Since the time complexity of extracting each gap is constant (each gap primitive has up to six edges), the total time complexity of the gap extraction algorithm is O$(n_t)$, where $n_t$ is the number of triangles. However, given an initial sampling generated by dart-throwing, the number of gap triangles is much less than the total number of triangles. We provide more details in Appendix~\ref{app:decom} to show the validness of this process.

\paragraph*{Boundary handling}
We also consider bounded domains. The boundary can be a simple polygon with or without holes. In these cases, we first compute the clipped power diagram using the technique presented in~\cite{Yan2012}. The gaps that touch the domain boundary are simply computed by clipping the power cells that contain parts of gaps with the surrounding disks. Figure~\ref{fig:adp2d} shows a result of sampling/meshing a 2D polygon.

\subsection{Gap primitive updates} \label{sec:update}

Appropriate algorithms exist to construct regular triangulations~\cite{Edelsbrunner1996}. The points can be dynamically inserted and deleted from a regular triangulation. Inserting points requires O($n^2$) in the worst case  and O(log($n$)) in the average case~\cite{Edelsbrunner1996}, where n is the total number of points. A point can be deleted in constant time~\cite{Devillers2006}. Deleting $k$ points requires O($k$) time.

If all points move in a coherent fashion, the regular triangulation might be updated efficiently by existing algorithms~\cite{Vigo2002}, but in our implementation, we simply rebuild the data structures if all points move at once. Movement of individual points is handled by deletion and insertion.

Changing the sampling radius of all disks causes a change in the combinatorial structure and we also have to rebuild the regular triangulation. However, if the sampling radius is constant, the regular triangulation is equivalent to the Delauney triangulation and remains unchanged in this case.

\section{Gap Computation on Surfaces} \label{sec:gapsurf}

\begin{figure}[ht]
\centerline{
\includegraphics[width=0.98\linewidth]{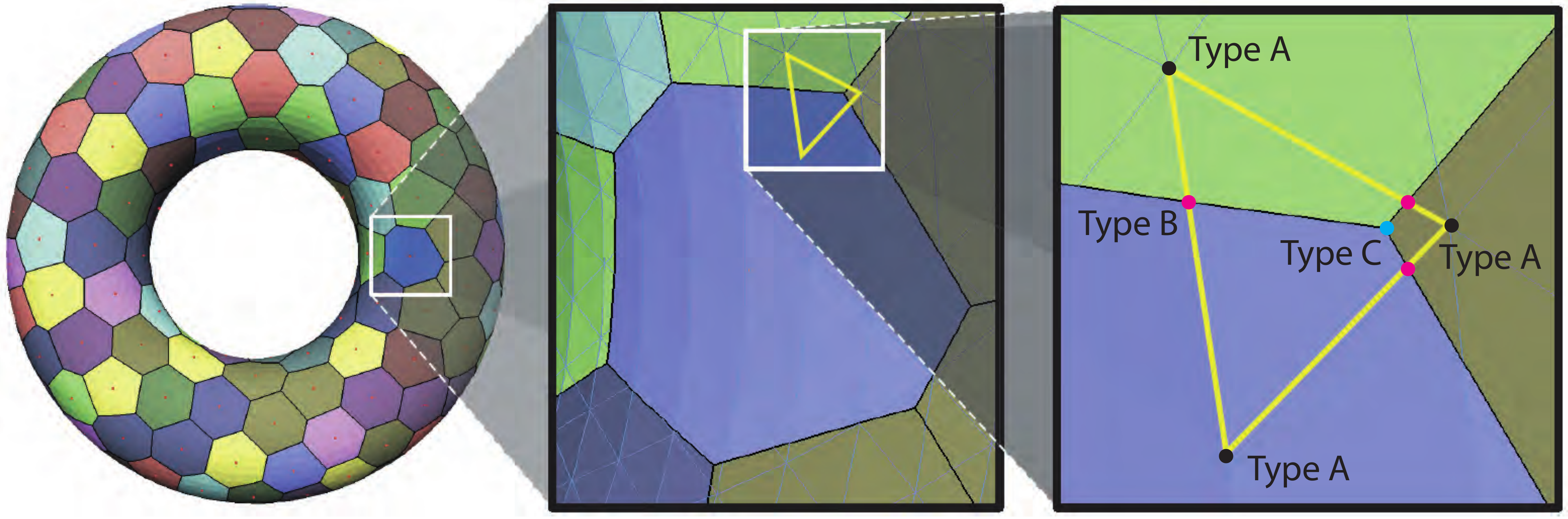}
}
\caption{Illustration of the restricted power diagram on a surface (left); middle: a restricted power cell; right: a mesh triangle is split into polygons shared by the incident cells.}
\label{fig:rpd}
\end{figure}

In this section, we generalize the gap analysis and computation framework to mesh surfaces. Suppose that the input domain $\Omega=\{f_j\}_{j=1}^m$ is a triangle mesh surface (where $f_j$ refers to a triangle) and $\wps$ is a set of weighted samples lying on the surface.
Recall that in Section~\ref{sec:gapexist}, we defined the restricted power diagram on surfaces. In this case, each restricted power cell is the intersection of the 3D power cell and the mesh surface, i.e., $\Omega_i\bigcap\Omega=\{\bigcup\{\Omega_i\bigcap\,f_j\}, \forall\,f_j\in\Omega\}$. Figure~\ref{fig:rpd} shows an example of the $\rpd$ of a torus. There are three types of vertices in the $\rpd$ of a mesh surface, called restricted power vertices, i.e., A) the original vertices of the mesh, B) the intersection of a mesh edge and a bi-sector power plane, and C) the intersection of a power edge and a mesh triangle (Figure~\ref{fig:rpd}(right)). Thm.~\ref{thm:gapsurf} can be adapted as follows:

\begin{theorem}
A gap of a disk set $\wps$ exists on a mesh surface iff there exists a restricted power cell of a sample $\mp_i$, whose restricted power vertices are not fully covered by the respective sphere $(\mp_i, r_i)$ centered at $\mp_i$ with radius $r_i$.
\label{thm:gapmesh}
\end{theorem}

\begin{figure}[ht]
\centerline{
\includegraphics[width=1.0\linewidth]{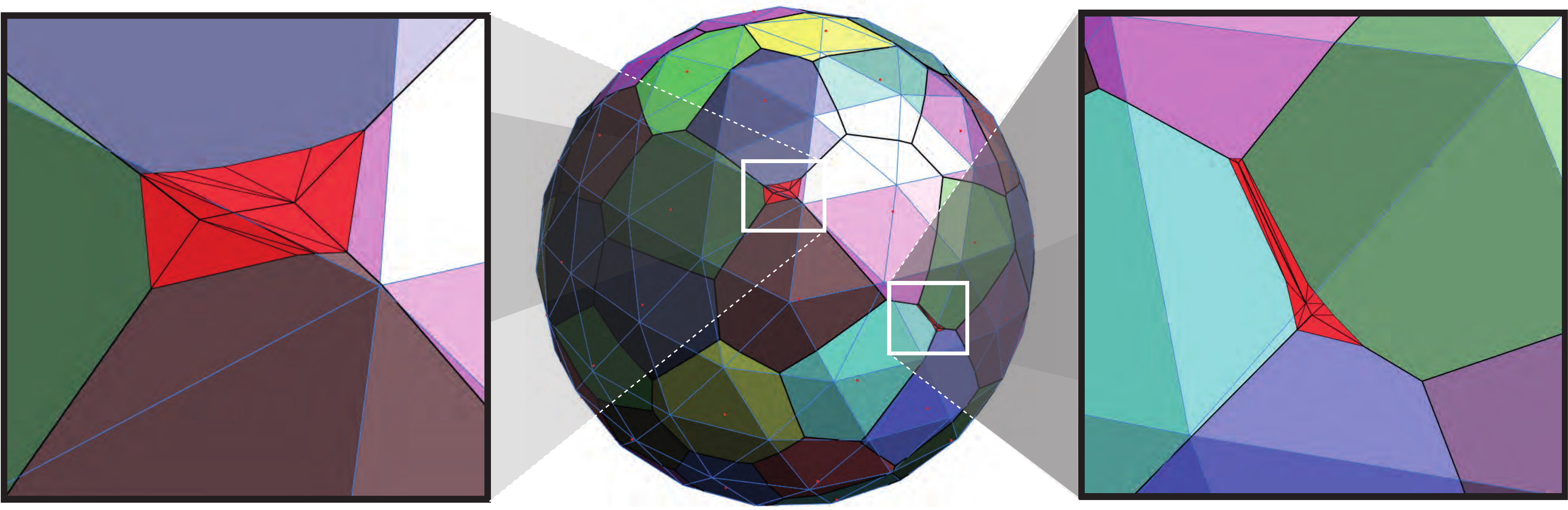}
}
\caption{Gap computation on surfaces by triangle-sphere clipping. Gaps are shown in red.}
\label{fig:surfgap}
\end{figure}

In the above definition, a type C restricted power vertex is also called a \emph{restricted power center}. Note that each restricted power center is dual to a triangle of the so-called restricted regular triangulation ($\rrt$). The triangles of the $\rrt$ do not lie on the surface, only the vertices do.
Since the restricted triangles no longer lie on the surface, we cannot use the 2D per-triangle gap computation (Section~\ref{sec:gap2d}) for meshes. Alternatively, based on Thm.~\ref{thm:gapmesh}, we present an approach similar to \cite{Jones2006} to compute the gap primitives on surfaces.
The gap primitives are computed by clipping the restricted power cell by the sphere centered at each weighted point. Each restricted power cell can be split into a set of triangles. Then, the clipping problem is reduced to a triangle-sphere intersection problem. The clipped regions, i.e., the gap primitives, are approximated by a set of triangles and associated with their incident gaps, as shown in Figure~\ref{fig:surfgap}.
The details of the $\rpd$ computation are analogous to the restricted Voronoi diagram computation in~\cite{Yan2009}.

Now we are able to identify and compute gaps on mesh surfaces. However, in the maximal sampling framework (see Section~\ref{sec:sampling}), we are willing to cluster the independent gap sets as in the 2D counterpart, which can be used for parallel gap filling. A simple region-growing-based approach can be used by detecting the connectivity between clipped gap primitives. Furthermore, if the initial sampling $\wps$ adapts to the local properties of the mesh surface (e.g., the \emph{$\epsilon$-sampling property}~\cite{Amenta1999}) and the \emph{topological ball property}~\cite{Edelsbrunner1997} is met, then the dual restricted regular triangulation is homeomorphic to the input domain $\Omega$. In this case, we are able to define the 3D gap triangles similarly to the 2D plane. A gap triangle is a triangle of the $\rrt$ with at least one vertex whose restricted power cell is not fully covered by the sphere centered at the vertex. Once the gap triangles are detected, we group the neighboring gap triangles/primitives into IGSs for further processing.


\section{Surface sampling and remeshing}
\label{sec:applications}

In this section, we first describe a framework for adaptive Poisson-disk sampling on surfaces, which also works well in 2D. Then, we present a brief discussion of the link between Poisson-disk sets and surface remeshing.  Moveover, we present a high-quality surface remeshing algorithm, as well as a randomized mesh optimization algorithm built on top of the sampling framework, which greatly improves the sampling/meshing quality.

\subsection{Adaptive sampling on surfaces} \label{sec:sampling}

As input, we use a mesh $\Omega=\{f_i\}_{i=1}^m$, a minimal sampling radius $r_{min}$, a maximal sampling radius $r_{max}$ (default value $16\,r_{min}$), as well as a density function $\rho(\mx)$ defined on the mesh surface. A voxel grid is built for accelerating the sampling process. We first voxelize the mesh with voxels whose sizes are equal to $\frac{r_{min}}{\sqrt{3}}$. Each voxel records the indices of samples that fully cover it. A voxel is \emph{valid} if it is not fully covered by any sphere. We follow the two-step sampling strategy presented in~\cite{Ebeida2011}: 1) dart throwing with a grid and 2) gap filling.

\paragraph*{Initial sampling}
In the first step, we perform classic dart throwing on surfaces. Each triangle $f$ is associated with a weight
$w_f=\rho(\mc_f)|f|$, where $\mc_f$ is the barycenter and $|f|$ is the area of the triangle. The cumulative probability density function (cpdf) of the weights is stored in a flat array. Each time, a dart is generated by first
selecting a triangle from the cpdf and then a point $\mp$ is randomly generated in the triangle, as well as a radius $r=\min(r_{max},\max(r_{min},\frac{1}{\sqrt{\rho(\mp)}}))$ associated with the point. The new dart is tested against the grid. The dart is accepted if it is not contained by previous samples and does not contain any other existing samples. The index of the new sample is recorded by the voxels that are fully contained by this sample. The first step is terminated when $k$ consecutive rejections are observed ($k=300$ in our implementation).

\paragraph*{Gap filling}
The second step is an iterative algorithm. At each iteration, all gap triangles are first grouped into IGSs. All gap primitives of each IGS are triangulated and each triangle is associated with a weight as before. For each IGS, we create a cpdf for the triangles belonging to it. Since the sampling of each IGS does not affect another IGS, we perform gap filling in parallel. We use the same method as in the initial sampling stage to generate new disks for each IGS. Note that in the gap filling step, the newly generated disks whose centers are not covered by existing disks may cover the centers of the existing disks. In this case, we recompute the radius of the new disk so that it will not contain any previous samples, i.e., we set the radius of the new disk as the distance
to the nearest sample. Here, we choose the maximal conflict metric~\cite{Kalantari2011} since we assume that the centers of the disks cannot be covered by any other disks. This solution is reasonable for a smooth density function. Actually, our gap computation framework can also handle the cases when the disk centers are covered by other disks. This will be further explored in future work. While there are multiple applications for sampling points on surfaces, the most important application is remeshing, which we will describe next.

\subsection{Poisson-disk sampling and surface remeshing} \label{sec:pdremesh}

Surface remeshing is a broad topic and the best choice of a surface remeshing algorithm depends on the application~\cite{Alliez2008}. Poisson-disk sampling is mainly useful for remeshing for simulations. In other applications, e.g., architectural panel layouts, the blue noise pattern will often be considered unattractive.  Before presenting our solution to adaptive remeshing, we would first like to discuss what criteria are required for good mesh and also establish the link between our disk sampling method and remeshing. This discussion suggests why our work can be more successful than previous surface sampling algorithms at remeshing for simulation.

First, the blue noise property is important because it reduces directional bias in the simulation~\cite{Ebeida2012}. While this fact has been used for uniform remeshing using disk sampling with a global uniform radius, it is also important for adaptive remeshing. Second, the most important geometric characteristic of a triangle mesh is the minimal angle in a triangle or the percentage of triangles with small angles~\cite{Shewchuk02}. Statistics relating to the minimal angle $\theta_{min}$, as well as $Q(t)=\frac{6}{\sqrt{3}}\,\frac{|t|}{{p}(t)\,{h}(t)}$, (where $|t|$ is the area of $t$, ${p}(t)$ is the half-perimeter of $t$ and ${h}(t)$ the longest edge length of $t$~\cite{Frey1997}) are mentioned in almost all recent remeshing papers. The reason for this is that the minimal angle influences the condition number of the matrices in FEM. Interestingly, there is a theoretical guarantee that the minimal angle is 30 degrees in a uniform 2D maximal sampling~\cite{Chew1989,Ebeida2011b}. It can also be shown that the same conclusion still holds on surfaces. However, in the case of adaptive sampling, we do not have any angle guarantees.
We propose a randomized optimization technique that can improve the angle bounds dramatically for adaptive remeshing (see Section~\ref{sec:remesh}).
This is an important link that is often overlooked in previous work: maximal sampling is essential for theoretical guarantees about triangle angles. Maximal sampling is also important in praxis. We verified this by generating $100$ point sets with $10k$ samples and then removing the last $20$ sampled points. The minimal angle is significantly worse. Only in about $10\%$ of the cases it is still $\geq 30^o$.
Third, irregular vertices are undesirable, because they require special handling and often separate code. Therefore, we aim at reducing the different types of irregular vertices to only two types: valence $5$ and valence $7$ vertices. It would also be helpful to reduce the total number of irregular vertices, but this conflicts with the blue noise property and the minimal angle properties.
Fourth, the surface approximation is important. Most commonly, this is measured in the Hausdorff distance or the root mean squared error.
Fifth, it is important to use Euclidean distances and not geodesic distances. While sampling geodesic disks seems more complex and sophisticated, this can be counterproductive for remeshing. The edges in a mesh are straight and not geodesic paths on surfaces. A simple implication is that the use of geodesic distances voids all angle guarantees.

\subsection{Surface remeshing and optimization} \label{sec:remesh}

To obtain a mesh, we extract the dual triangulation from the restricted power diagram of the samples, which is a good remeshing of the input surface $\Omega$. There are several potential challenges to surface remeshing that we address here.

\paragraph*{Topological validity}
In this work, we adapt the sampling radius to the local feature size to ensure the topological validity. Theoretically, the $\epsilon$-sampling theorem~\cite{Amenta1998} requires that for each point $\mx\in\Omega$, there exists a sample $\mp$, such that $|\mx-\mp|<0.3\,lfs(\mp)$.  The $\rrt$ of the samples is then homeomorphic to $\Omega$. However, in practice, even if these theoretical guarantees are not met, our algorithm still generates topologically correct results as shown in Figure~\ref{fig:remesh}. We can detect topological inconsistencies using the algorithm presented in~\cite{Yan2009}. Once such a configuration is detected, we discard the samples that cause the problem and resample the gaps. We refer to~\cite{Yan2009} for more details of the mesh extraction algorithm since it is not the main contribution of this paper.

\paragraph*{Valence and angle optimization}
Uniform blue noise remeshing has many nice properties. Given a constant sampling radius $r$, the meshes generated from blue noise sampling exhibit the following bounds: an angle bound $[30^o,120^o]$, an edge length bound $[r,2\,r]$ and an area bound $[\frac{\sqrt{3}}{4}\,r^2,\frac{3\,\sqrt{3}}{4}\,r^2]$ as its 2D counterpart. All these properties are desired in many applications, particularly the angle bound, which is crucial to FEM applications~\cite{Shewchuk02}. However, in the case of adaptive sampling, the above theoretical bounds do not hold. To improve the meshing quality, we introduce a novel and simple randomized optimization algorithm, i.e., angle bound and valence optimization.

This algorithm iteratively removes the sample points with unsatisfactory properties and their neighborhoods and then refills the gaps. In angle bound optimization, the vertices with one triangle angle less than a minimal angle threshold or larger than a maximal threshold are removed. In valence optimization, the vertices whose degrees are less than 5 or larger than 7 are removed. These two optimization criteria can either be performed separately or jointly. The valence/angle optimization terminates when the required criteria are met or the maximal iteration number (25 in our implementation) is reached. In a joint valence and angle optimization, a global optimization is performed interleaving between valence and angle optimization. Typically, it takes 5-10 global iterations to meet both quality requirements. During the optimization, the $\rrt$ and $\rpd$ are locally updated.

\paragraph*{Edge length optimization}
Besides the valence/angle optimization, we are also able to optimize other geometric properties, such as edge length, using the same framework.  Given a user input threshold $\lambda_e$, we iteratively remove all the edges $(\mp_i, \mp_j)$ with $|\mp_j-\mp_i|>\lambda_e\,(r_i+r_j)$, followed by a gap filling step, until the desired edge length quality is met.

\paragraph*{Feature preservation}
We also implemented a simple feature-preserving sampling step in our framework. The features are provided by the user, in the form of a 1D curved skeleton.
The corners (vertices of the skeleton with more than 2 neighboring edges, or with sharp turning angles) are inserted directly as sample points, and the feature curves are first sampled before the surface sampling. The edge lengths of the feature samples are optimized by edge length optimization ($\lambda_e=\sqrt{3}$) so that the neighboring spheres are deeply intersected~\cite{Cheng07b}. The samples of the feature skeleton remain fixed during the surface sampling/optimization stages.

\section{Results and Discussion}

In this section, we present several sampling and meshing results using our framework based on the proposed gap processing techniques. We also compare our results with current state-of-the-art approaches. We use CGAL~\protect\cite{cgal} for the sequential 2D/3D regular triangulation, and the OpenMP library for parallel gap filling. The experimental results are conducted on an Intel X5680 Dual Core 3.33GHz CPU with 4GB memory and a 64-bit Windows 7 operating system.

\subsection{2D sampling and meshing}

\begin{figure}[t]
\centerline{\includegraphics[width=1.0\linewidth]{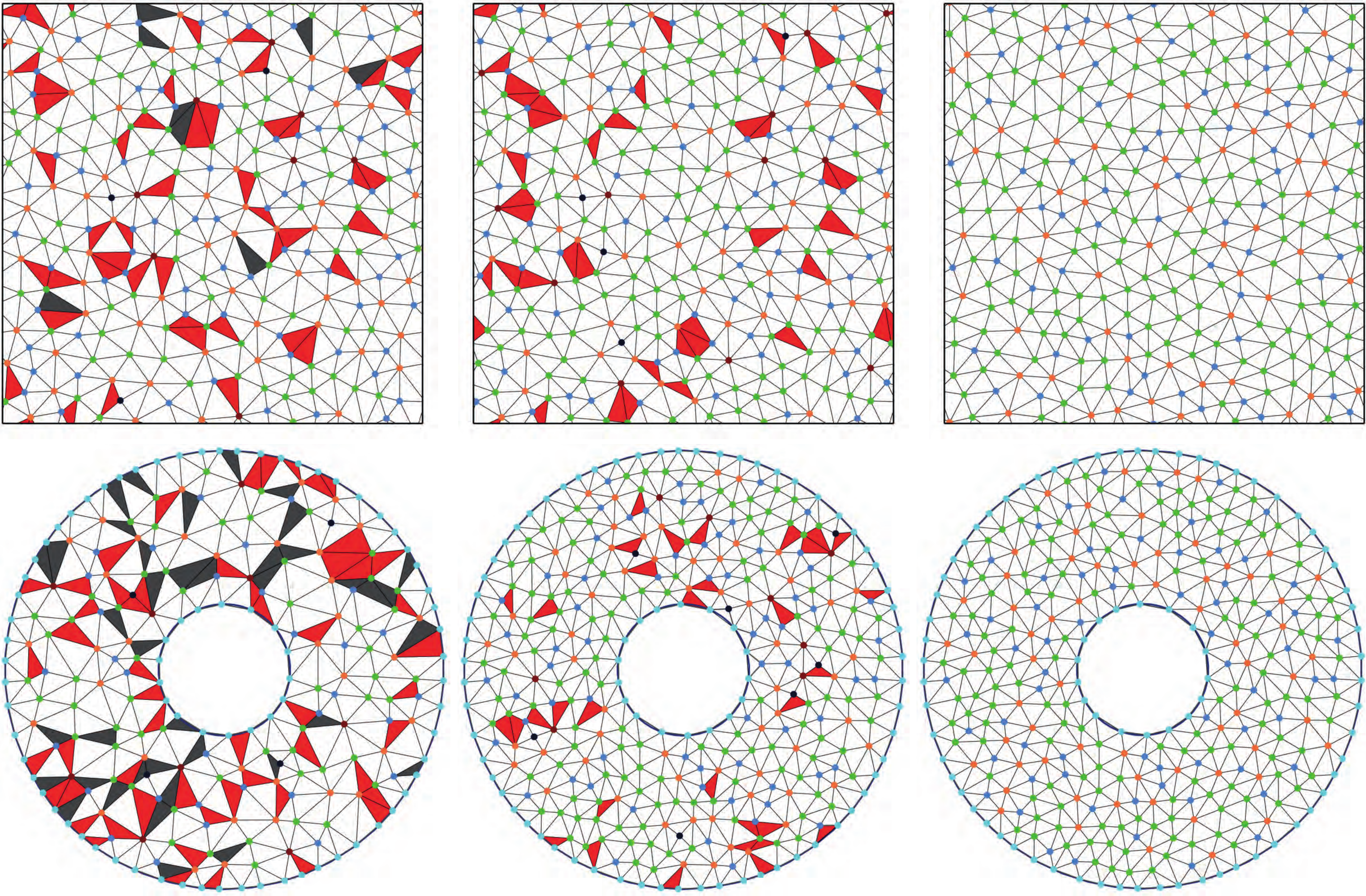}}
\caption{Illustrations of uniform 2D sampling/optimization. Top: a unit square with a periodic boundary condition; bottom: two concentric circles. Left column: dart throwing; middle: maximal sampling; right: optimized sampling (angle and valence). Vertices with valence 5: blue, 6: green, 7: orange. Darker points correspond to higher ($>7$) or lower ($<5$) valences. The triangles with $\theta_{min}<30^o$ or $\theta_{max}>120^o$ are shown in dark gray, triangles with $\theta_{min}\in[30^o,35^o]$ or  $\theta_{max}\in[105^o,120^o]$ are shown in red.}
\label{fig:uniform2d}
\end{figure}

We first present the results of 2D sampling. Figure~\ref{fig:uniform2d} shows two examples of uniform sampling and optimization. The first example is a 2D square with a periodic boundary condition, and the second is defined by two concentric circles. The statistics of these results are given in Table~\ref{tab:uniform2d}.
\begin{table*}[!htp]
\centering
{
\begin{tabular*}{0.9\linewidth}{@{\extracolsep{\fill}}|c|c|c|c|c|c|c|c|c|c|}
\hline Result & \#v & $\theta_{min}$ & $\theta_{max}$ &$|e|_{min}'$ & $|e|_{max}'$ & $|t|_{min}'$ & $|t|_{max}'$ &$\theta<30^o$ & $v_{567}$ \\
\hline Square1  & 26.8k  & 21.6  & 130.3 & 1.0  & 1.458 & 0.947 & 2.043 & 3.33 & 94.2\% \\
\hline Square2  & 34.5k  & 30.1  & 118.6 & 1.0  & 0.999 & 1.001 & 0.998 & 0.00 & 96.4\%  \\
\hline Square3  & 35.1k  & 35.0  & 104.9 & 1.0  & 0.999 & 1.001 & 0.997 & 0.00 & 100\%  \\
\hline Circles1  & 19.6k  & 18.9  & 131.4 & 1.0  & 1.950 & 1.088 & 2.950 & 8.83 & 93.1\%  \\
\hline Circles2  & 24.9k  & 30.0  & 118.5 & 1.0  & 0.999 & 1.003 & 0.996 & 0.00 & 96.2\%   \\
\hline Circles3  & 25.3k  & 35.0  & 105.0 & 1.0  & 0.999 & 1.003 & 0.995 & 0.00 & 100\%   \\
\hline
\end{tabular*}
} \caption{Statistics of 2D uniform sampling/meshing ($r_{min}=4.5\times\,10^{-3})$.
$|e|_{min}' = \frac{|e|_{min}}{r_{min}}$ is the ratio of the minimal edge length over the theoretical minimal edge length bound, as it is for $|e|_{max}' = \frac{|e|_{max}}{2\,r_{min}}$, $|t|_{min}' = \frac{|t|_{min}}{\frac{\sqrt{3}}{4}\,r^2_{min}}$, and $|t|_{max}' = \frac{|t|_{max}}{\frac{3\,\sqrt{3}}{4}\,r^2_{min}}$. $\theta<30^o$ is the percentage of the triangles with $\theta_{min}$ smaller than 30 degrees. $v_{567}$ is the percentage of vertices with valence 5, 6, or 7. For each model, we show the meshing quality of 1) classic dart throwing, 2) maximal sampling and 3) optimized sampling. The data shown in the table are averages of 10 runs for each model.}
\label{tab:uniform2d}
\end{table*}

\begin{figure}[!htp]
\centering
\includegraphics[width=1.0\linewidth]{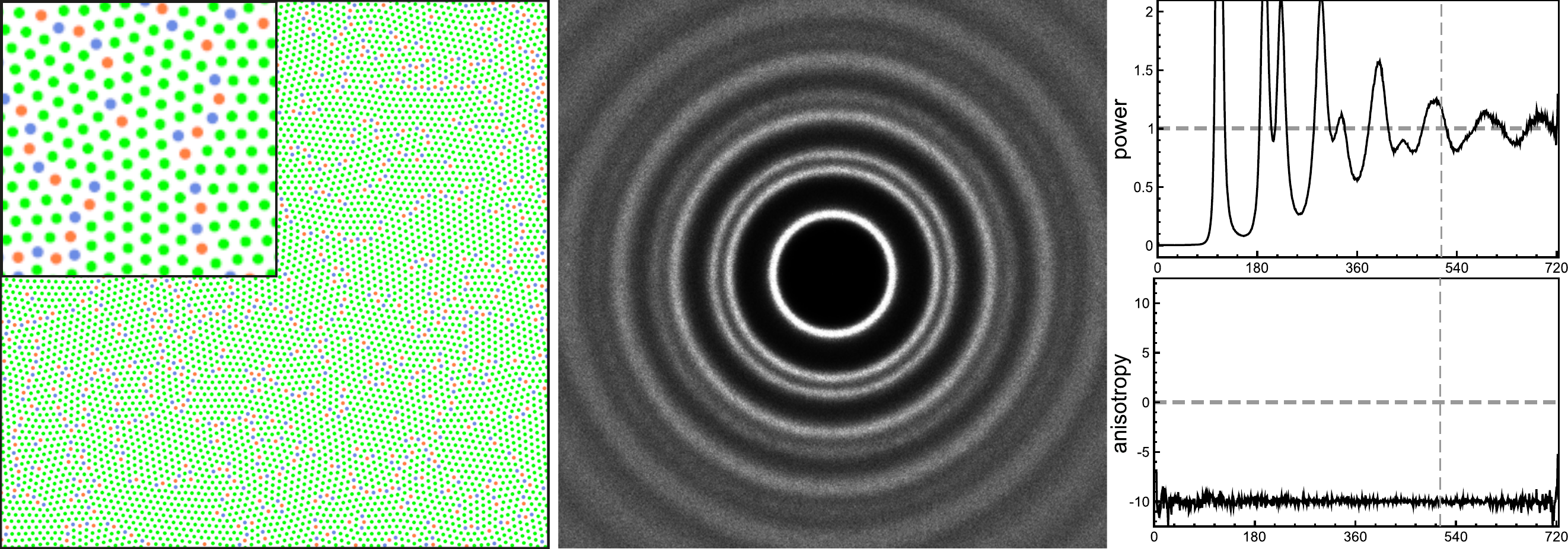}\\
\includegraphics[width=1.0\linewidth]{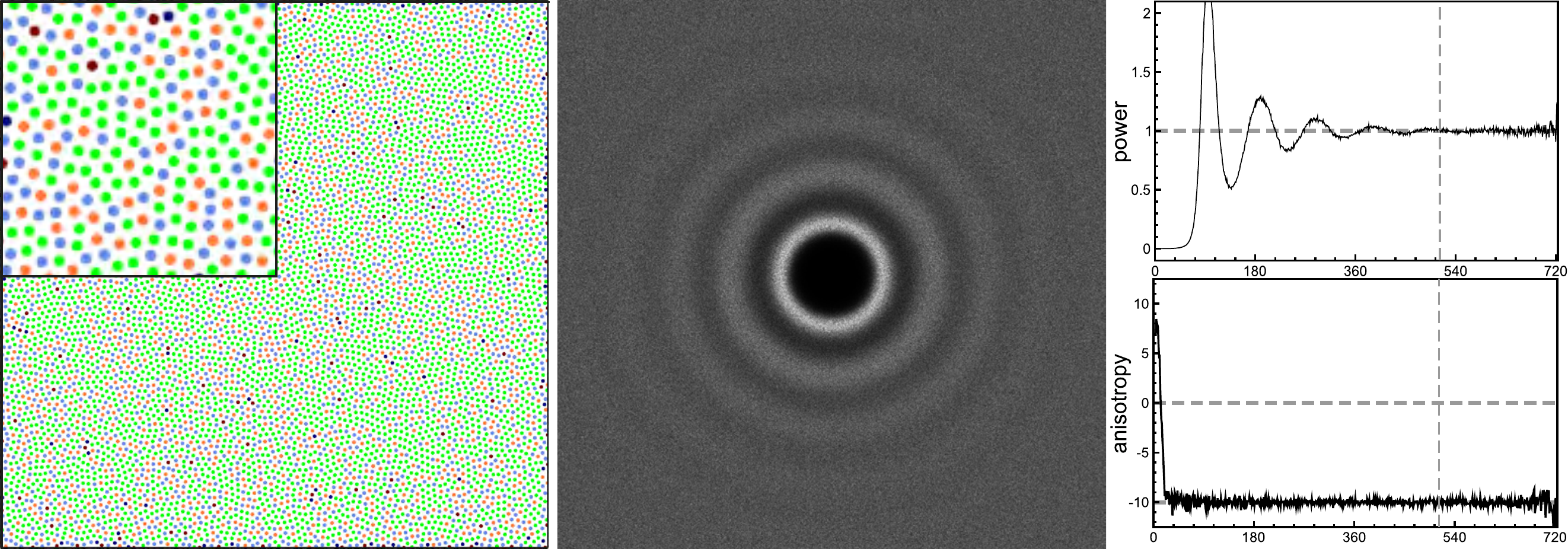}\\
\includegraphics[width=1.0\linewidth]{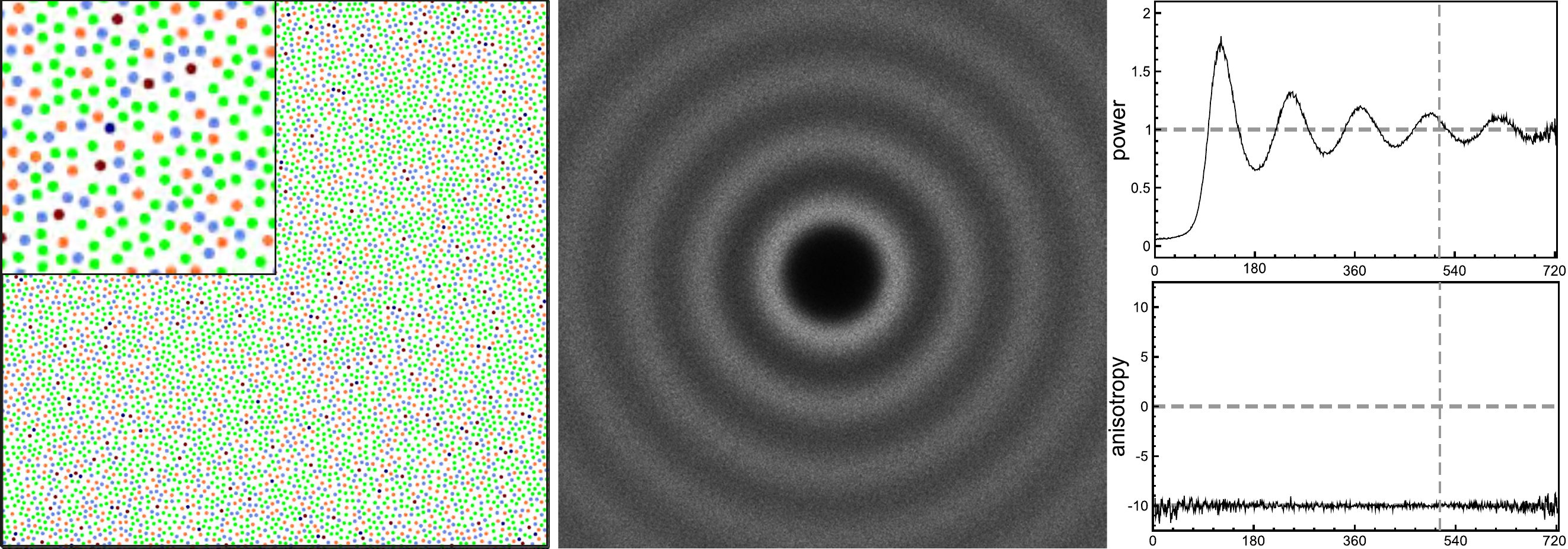}\\
\includegraphics[width=1.0\linewidth]{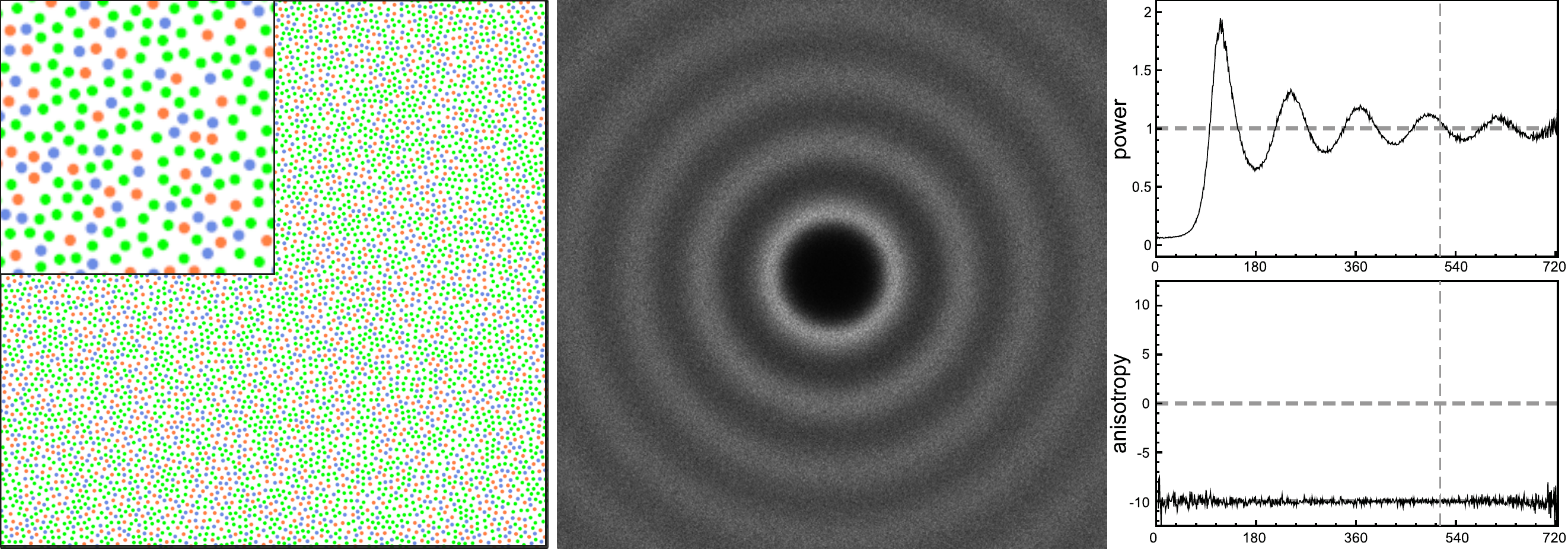}\\
\includegraphics[width=1.0\linewidth]{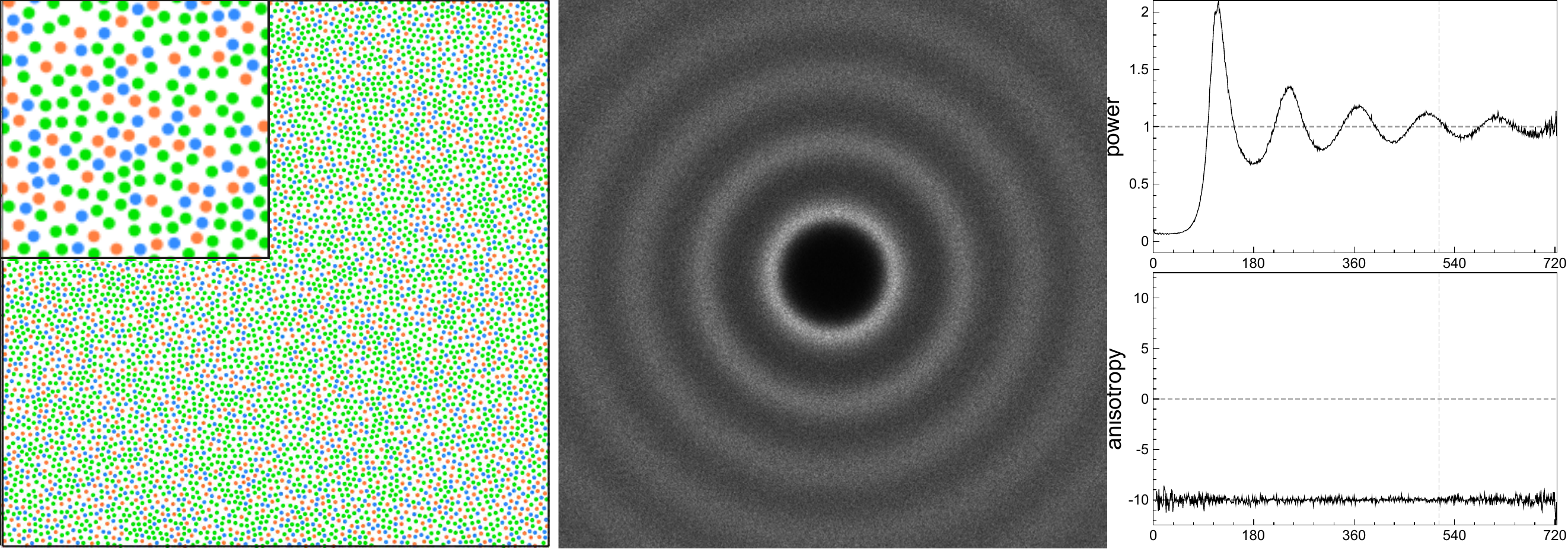}\\
\caption{ Spectral analysis of different approaches (10k samples). From top to bottom: Lloyd iteration~\protect\cite{Lloyd1982}; CCVT~\protect\cite{Balzer2009}; maximal Poisson sampling; valence optimization; valence and angle optimization. The left column is the point set generated by each method, where green dots show vertices with valence 6, red for valence 7, and blue for valence 5. Darker colors signify higher valences.}
\label{fig:spec}
\end{figure}

We show a comparison to uniform 2D sampling and the corresponding spectral analysis in Figure~\ref{fig:spec}, which demonstrates that our optimization framework preserves the blue noise properties well. We use the PSA software~\cite{Schlomer2011} to perform the spectral analysis of the Poisson-disk sets. Table.~\ref{tab:val} lists the statistics of vertex valences of different approaches.

\begin{table}[!htp]
\centering
{\scriptsize
\begin{tabular*}{1.0\linewidth}{@{\extracolsep{\fill}}|c|c|c|c|c|c|c|c|}
\hline           & $v4$  & $v5$ & $v6$  & $7v$  & $v8$ & $v9$ & $v10$\\
\hline Lloyd     & 0    & 7.42  & 85.15 & 7.43  & 0    & 0    & 0   \\
\hline CCVT      & 0.49 & 21.63 & 56.30 & 20.55 & 1.02 & $4.9\,e^{-5}$ & 0  \\
\hline MPS       & 1.23 & 24.04 & 50.48 & 22.06 & 2.18 & $4.7e^{-4}$ & $7.5\,e^{-7}$   \\
\hline Ours1     & 0    & 23.02 & 53.96 & 23.02 & 0    & 0    & 0   \\
\hline Ours2     & 0    & 22.01 & 55.98 & 22.01 & 0    & 0    & 0   \\
\hline
\end{tabular*}
} \caption{Statistics of the vertex valences of different methods. Each column is the percentage  of
the vertices of the corresponding valence. MPS stands for any maximal Poisson sampling generated by dart throwing. Ours1 is the result of valence optimization, and Ours2 is the valence and angle optimization.}
\label{tab:val}
\end{table}

\begin{figure}[ht]
\centerline {
\includegraphics[width=1.0\linewidth]{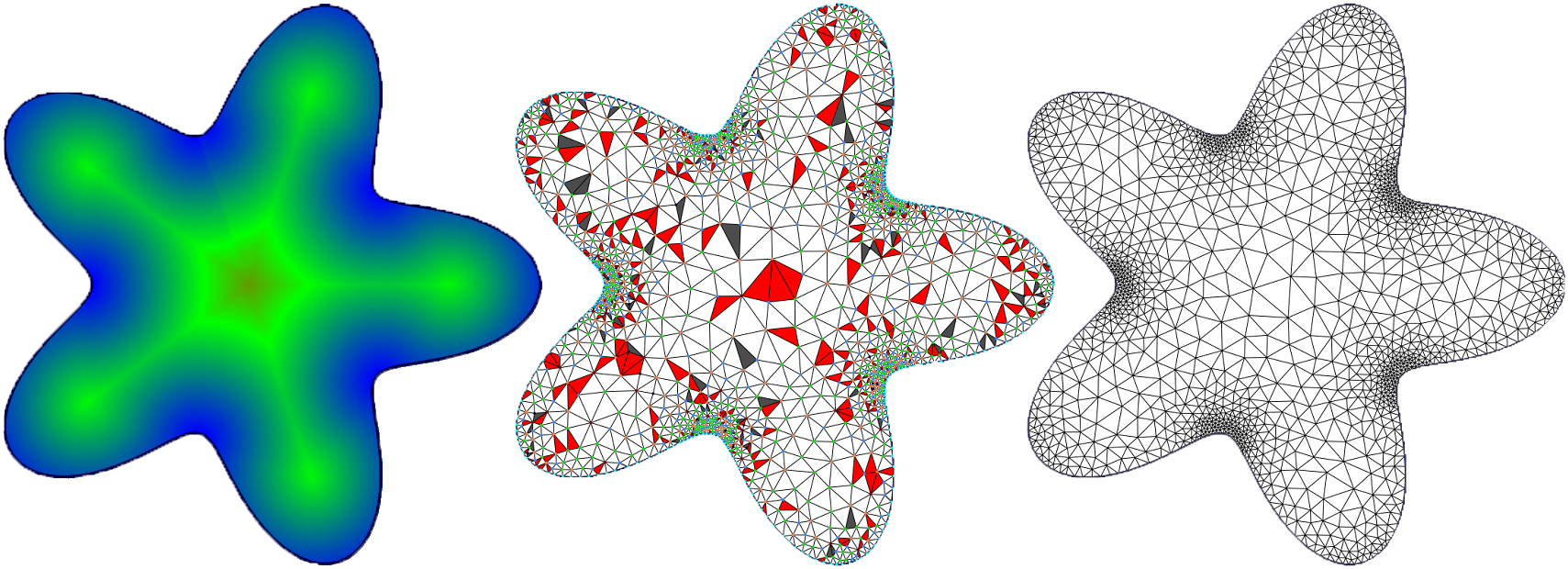}}
\caption{Adaptive 2D sampling/meshing. Left: density field; middle: initial sampling. Triangles with $\theta_{min}<30^o$ and $\theta_{max}>120^o$ are shown in dark gray and triangles with $\theta_{min}\in[30^o,35^o]$ and $\theta_{max}\in[105^o,120^o]$ are shown in red; right: optimized sampling.}
\label{fig:adp2d}
\end{figure}

An example of adaptive meshing is given in Figure~\ref{fig:adp2d}, where we set the desired angle bounds to $[35^o, 105^o]$. For the examples with boundaries, we first perform 1D maximal sampling on the boundary; then, we apply edge length optimization to ensure that the boundary disks are deeply intersected~\cite{Cheng07b} (we set $\lambda_e=\sqrt{3}$ in all our experiments). The boundary samples remain fixed during the later sampling/optimization inside the boundary.

\subsection{Comparison of 2D uniform MPS}
If the sampling radius is a constant $r$, then the regular triangulation/power diagram is equivalent to the Delaunay triangulation/Voronoi diagram. In this case, we can use the more efficient implementation of the Delaunay triangulation instead of the regular triangulation. In the following, we compare results from our framework with previous approaches for 2D uniform MPS. We call these approaches Ebeida~\cite{Ebeida2011b}, White~\cite{White2007}, Gamito~\cite{Gamito2009}, and Jones*~\shortcite{Jones2006}. Here, Jones* stands for an improved version of Jones' method \shortcite{Jones2006} where we changed the code to follow the framework of Ebeida by using a lazy update instead of continuously updating the data structures after each new sample.

\begin{figure}
\centerline{
\includegraphics[width=0.5\linewidth]{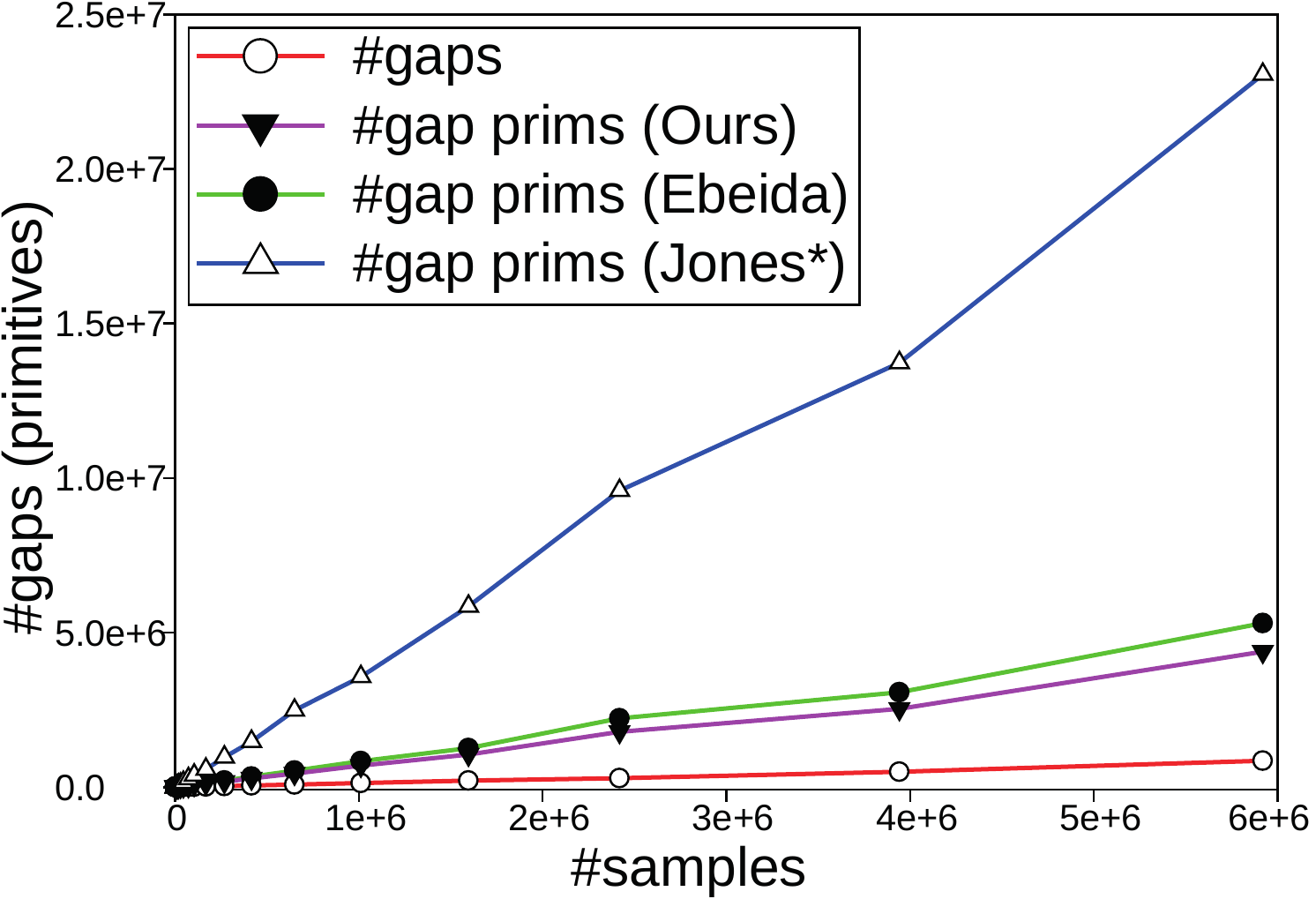}
\includegraphics[width=0.5\linewidth]{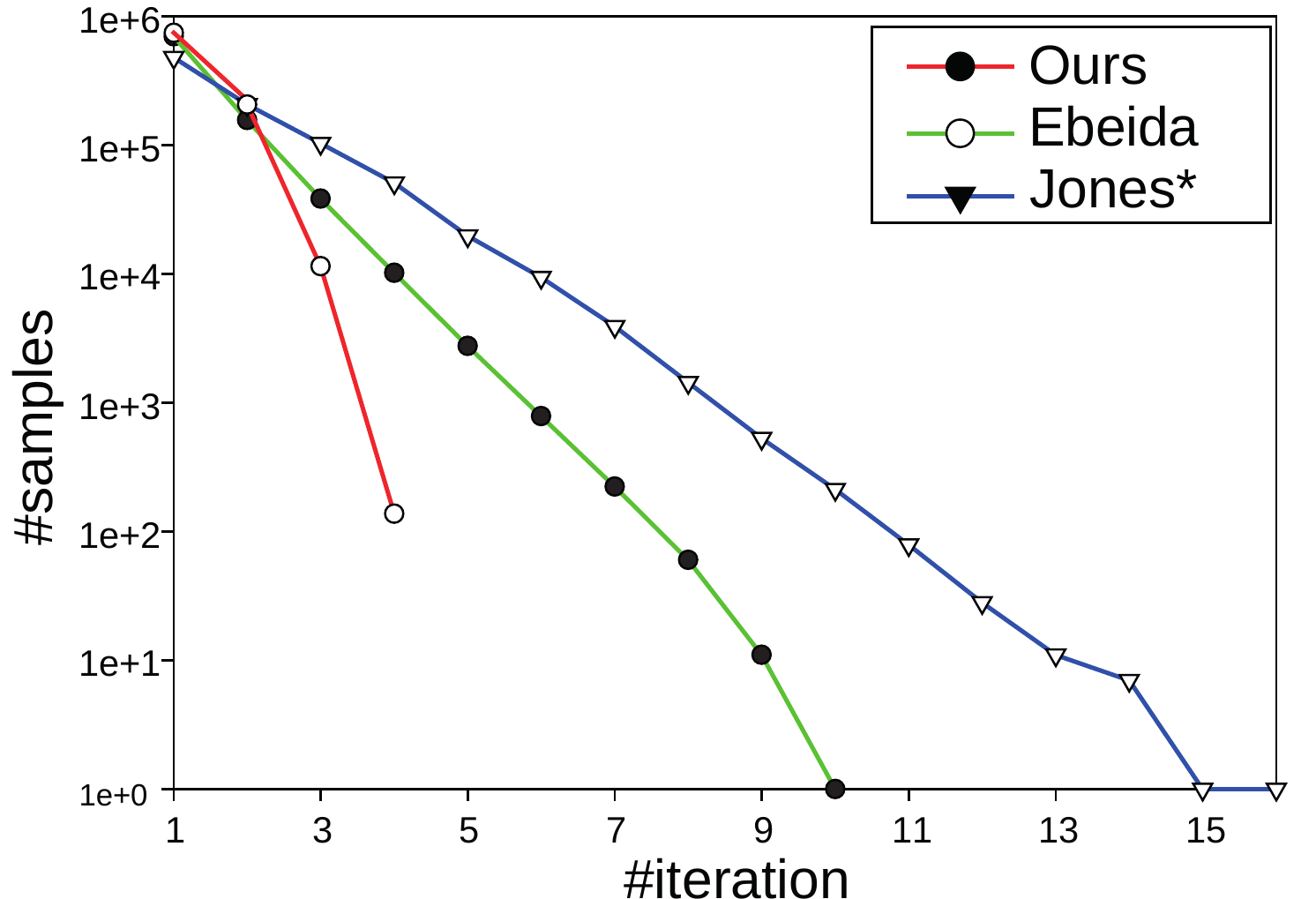}
}
\caption{The convergence of sampling independent gaps is much faster than sampling all gap primitives together. Left: comparison of the number of gap primitives. Right: our independent gap
sampling uses fewer iterations than do Ebeida and Jones*.}
\label{fig:converge}
\end{figure}

First, we show that our gap primitives capture the geometry of the problem very well.  Figure~\ref{fig:converge} (left) compares the number of gaps and gap primitives computed for a varying number of initial samples in the first stage. For 1M points, we obtain $129.3K$ gaps and $695.3K$ gap primitives compared with $841.7K$ gap primitives in Ebeida and $3.57M$ in Jones*.
In Figure~\ref{fig:converge} (right), we illustrate the advantage of extracting connected components. We compare the number of iterations in the second stage to generate a point set with sampling radius $r=5\times\,10^{-4}$ in a unit square domain with a periodic boundary (this gives approximately $2.8M$ points for each method). We perform multiple tests and select a representative experiment for each method. Typically our approach only needs 4 or 5 iterations to obtain a maximal point set while others need 10 or more iterations.

Second, we compare the running times of different algorithms. Our running time is slightly faster than the performance reported by White, Gamito, and Ebeida. However, we could only verify White's performance on our machine.
One advantage of our method compared with Ebeida is that it relies on Delaunay triangulation, a well-tested method where the code is publicly available, e.g., CGAL~\cite{cgal} and Qhull~\cite{Barber1996}\footnote{\texttt{http://www.qhull.org}}.  It takes only $1.91s$ to triangulate one million points obtained in stage one of the algorithm. Another advantage of Delaunay triangulation is that it can nicely track changes to the point set. For example, after an initial Delaunay triangulation is available, newly sampled points can be inserted into Delaunay triangulation using only O(1) instead of O(log($n$)), because we know the gap triangle that contains the new sampling point. This tracking ability is also especially useful for the applications we will describe next.

\begin{figure}[t]
\includegraphics[width=1.0\linewidth]{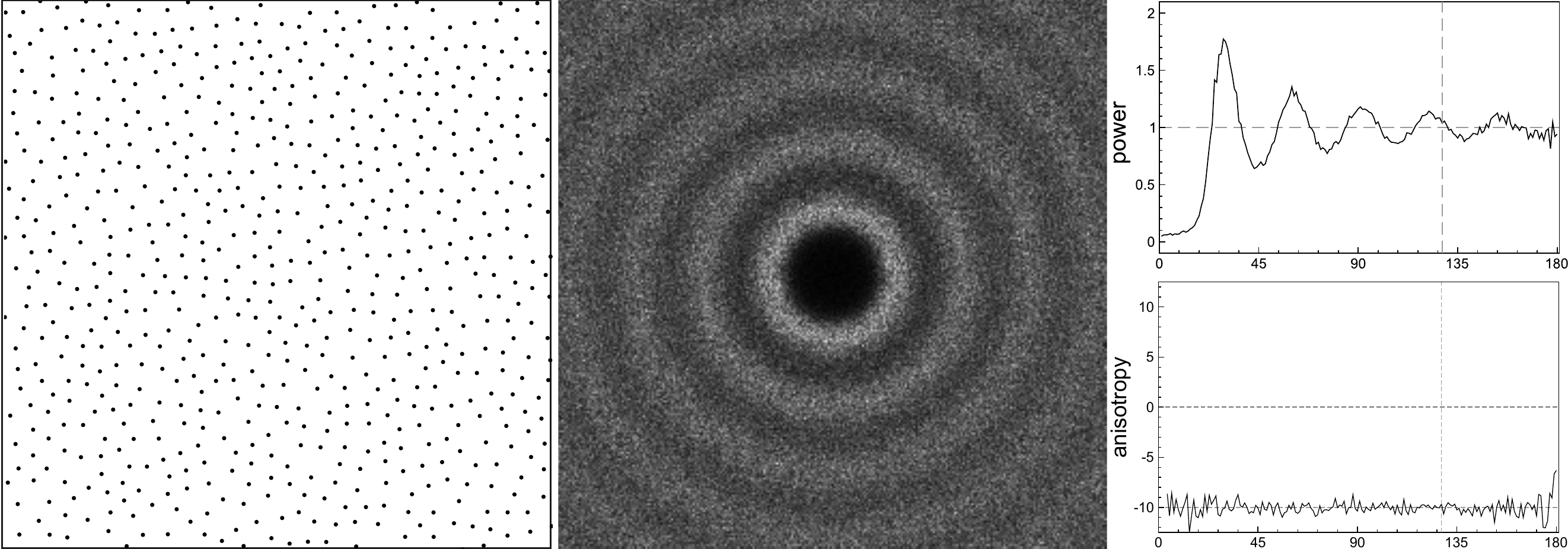}
\includegraphics[width=1.0\linewidth]{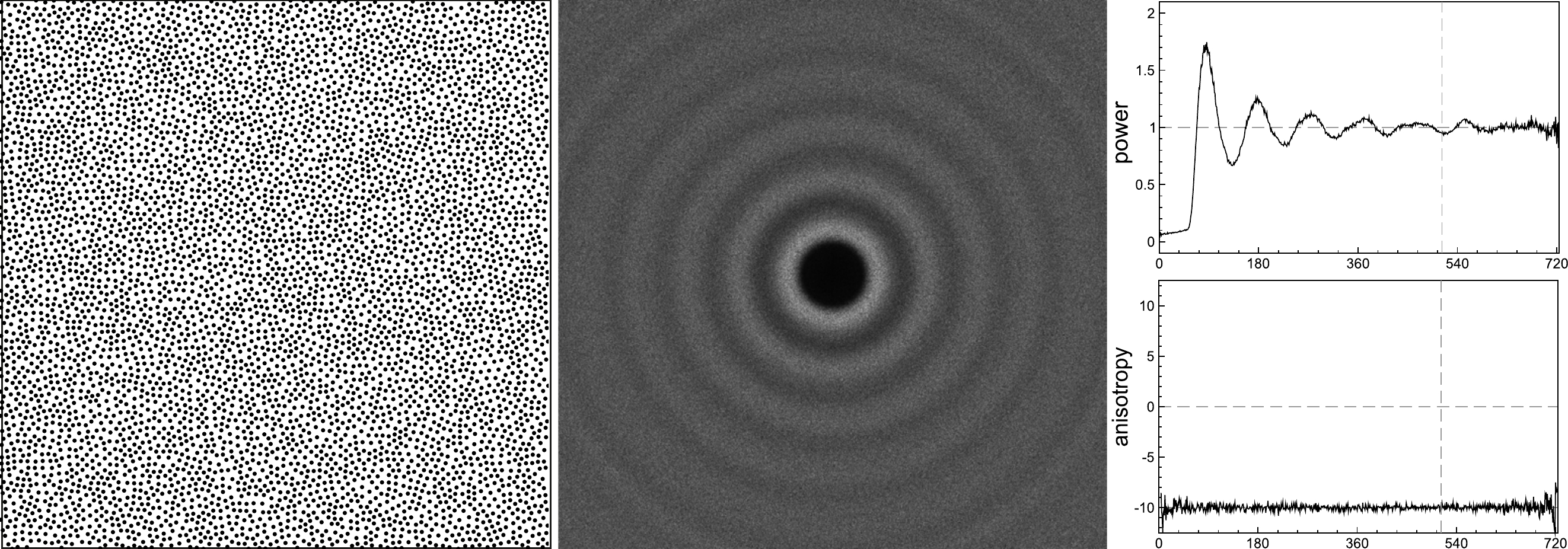}
\includegraphics[width=1.0\linewidth]{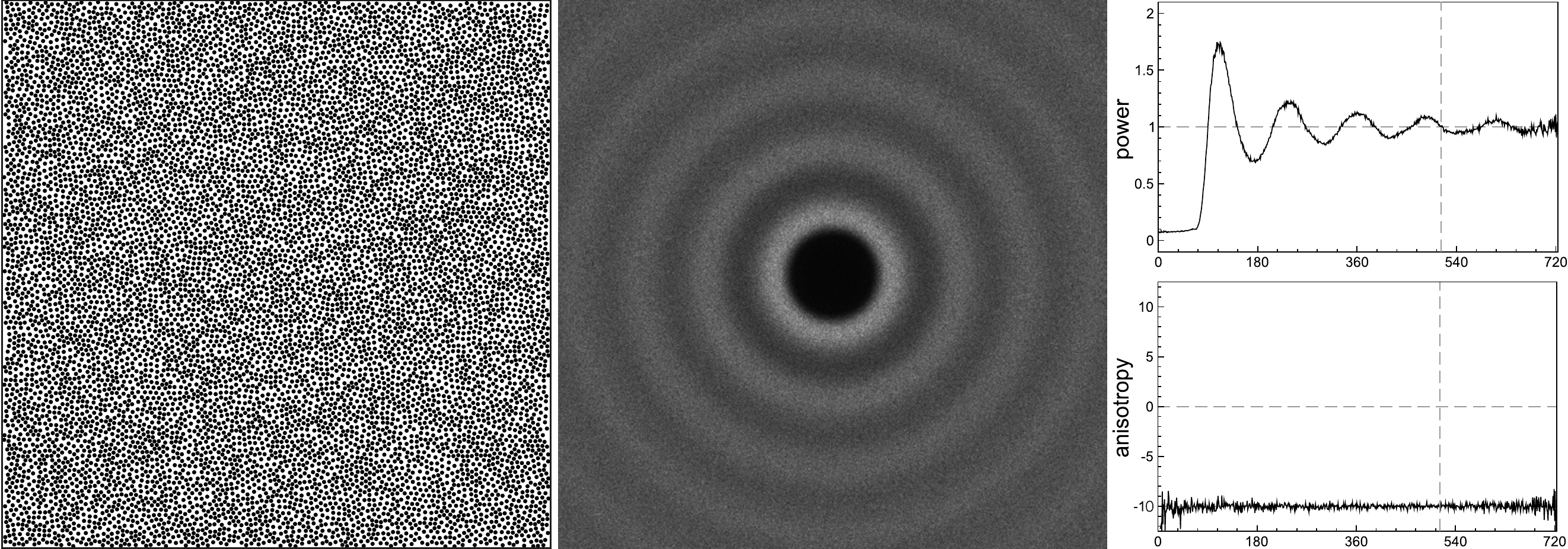}
\caption{Spectral analysis of different levels of hierarchical sampling. Top 680 points, middle 5.9K points and bottom 10K points. }
\vskip -0.1cm
\label{fig:hier}
\end{figure}

A good extension to dart throwing is to make the algorithm progressive by starting with a large radius and then continually shrinking the radius~\cite{Mccool1992}. In this way, a progressive randomized triangulation (or point set) is obtained where the final triangulation as well as all intermediate ones has nice spectral properties. We reimplemented~\cite{Mccool1992} and experimented with various heuristics to determine when to change the sampling radius to obtain a progressive triangle mesh (Figure~\ref{fig:hier}). In this application, we need the updating operation that changes the sampling radius.

\subsection{Surface sampling and remeshing}

\begin{table*}[!htp]
\centering
{
\begin{tabular*}{0.9\linewidth}{@{\extracolsep{\fill}}|c|c|c|c|c|c|c|c|c|c|}
\hline Result & \#v & $\theta_{min}$ & $\theta_{max}$ &$|e|_{min}'$ & $|e|_{max}'$ & $|t|_{min}'$ & $|t|_{max}'$ &$\theta<30^o$ & $v_{567}$ \\
\hline Sphere1 & 23.5k  & 22.1  & 129.3 & 1.0 & 1.618 & 0.993 & 2.182 & 1.12 & 94.5  \\
\hline Sphere2 & 29.7k  & 30.2  & 119.0 & 1.0 & 0.998 & 1.002 & 0.991 & 0.00 & 96.6  \\
\hline Sphere3 & 30.4k  & 35.0  & 105.0 & 1.0 & 0.998 & 1.002 & 0.996 & 0.00 & 100   \\
\hline Torus1  & 22.4k  & 21.3  & 126.8 & 1.0 & 1.382 & 0.979 & 1.816 & 0.88 & 94.6  \\
\hline Torus2  & 25.1k  & 30.1  & 119.1 & 1.0 & 0.998 & 1.004 & 0.994 & 0.00 & 96.3  \\
\hline Torus3  & 27.4k  & 35.1  & 104.9 & 1.0 & 0.999 & 1.003 & 0.995 & 0.00 & 100   \\
\hline
\end{tabular*}
} \caption{Statistics of uniform sampling/remeshing on surfaces ($r_{min}=4.5\times\,10^{-3}$). }
\label{tab:uniform3d}
\end{table*}

\begin{figure}[!htp]
\centerline{\includegraphics[width=1.0\linewidth]{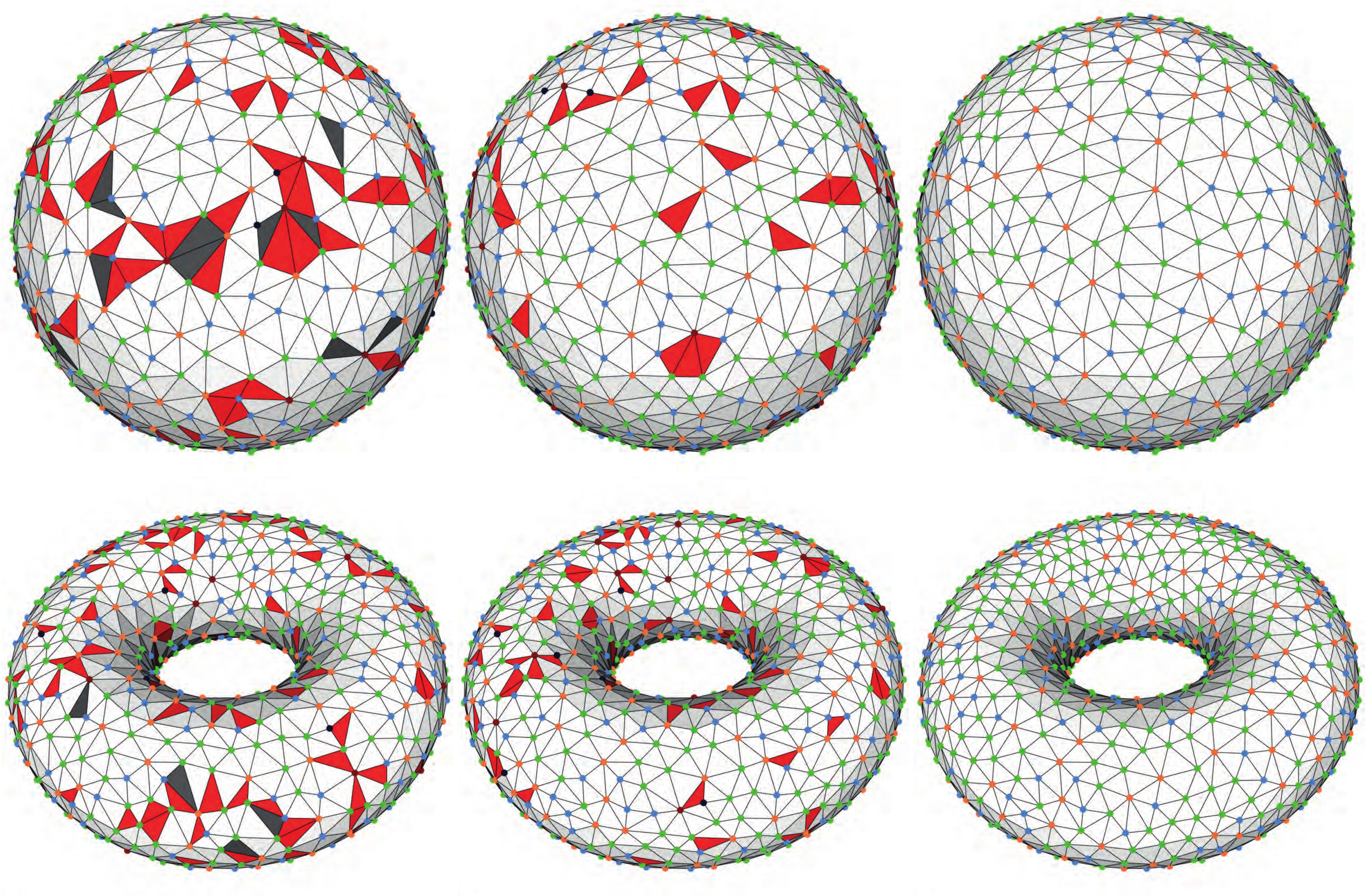}}
\caption{Illustration of uniform sampling/optimization on surfaces (we set $r_{min}=0.03$ for better visualization). Left column: dart throwing; middle: maximal sampling; right: optimized sampling (angle and valence). (Please refer to Figure~\protect{\ref{fig:adp2d}} for the color coding of vertices and triangles.)}
\label{fig:uniform}
\end{figure}

\paragraph*{Uniform sampling/remeshing}
We show several experimental results of surface remeshing and optimization. Figure~\ref{fig:uniform} shows the results of uniform sampling on a sphere and a torus. The minimal radius is set to $r_{min}=4.5\times{10^{-3}}$. In this example, we generate three meshes for each model using non-maximal sampling (dart throwing without gap filling), maximal sampling, and optimized sampling. The statistics of the meshes are shown in Table~\ref{tab:uniform3d}. We can see that the theoretical bounds do not hold for non-maximal sampling (first experiment), while these bounds are guaranteed by maximal sampling (second experiment). For optimized (uniform) sampling, the desired angle bound is set to $[35^o,105^o]$. This third experiment shows that the geometric properties are greatly improved.

\begin{figure}[!hpt]
\centerline{
\includegraphics[width=0.9\linewidth]{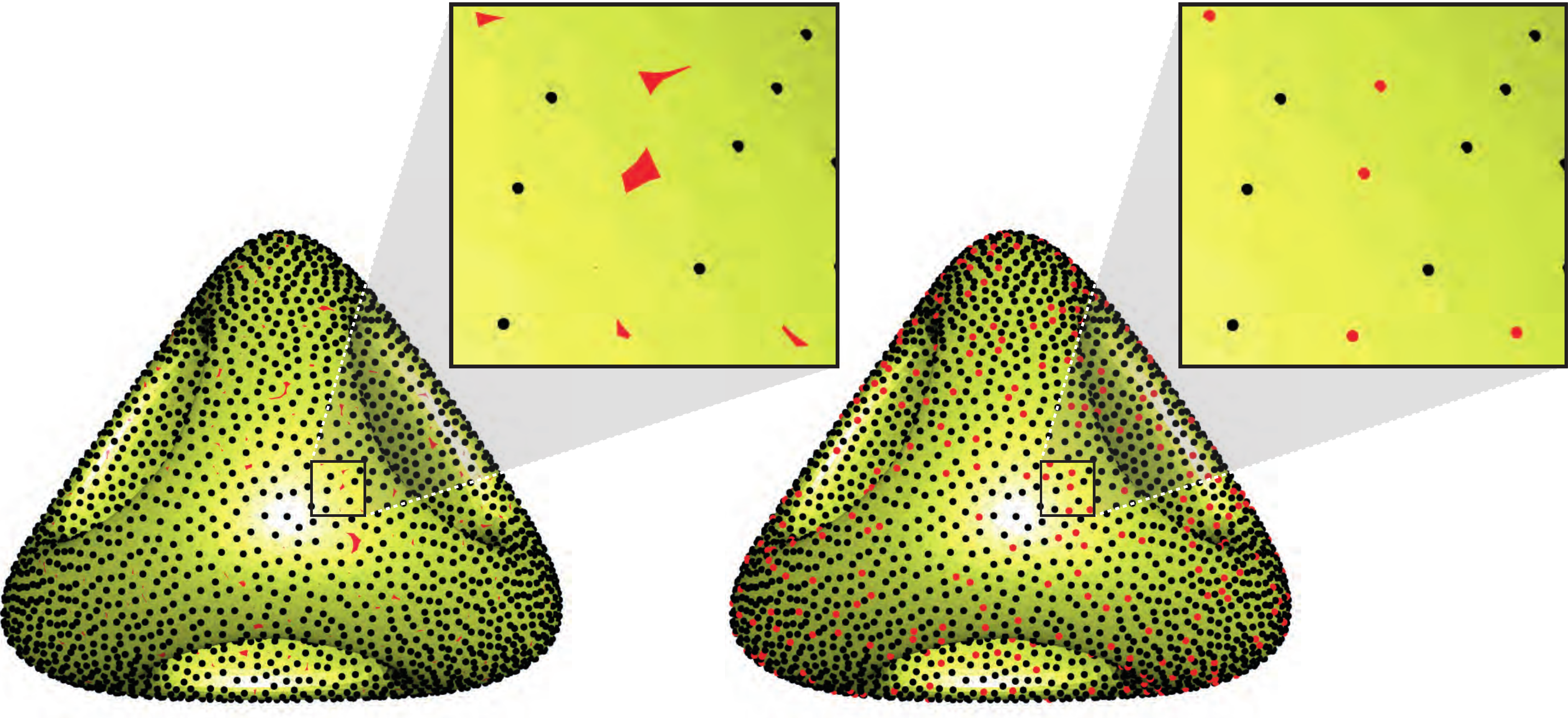}
}
\caption{Comparison with \protect\cite{Corsini2012}. Left: uniform sampling result of \protect\cite{Corsini2012} with 3909 samples ($r\approx\,0.012$). Gaps are detected by our technique (red regions); right: maximal sampling (4560 samples) by filling the gaps using our approach, and the new sampled points are shown in red (the model is from the Aim@Shape repository).}
\label{fig:cmp}
\end{figure}

We compare the uniform surface sampling with the most recent approach~\cite{Corsini2012}. To make a fair comparison, we set the over-sampling factor to $200$ so that their result has better maximality and the algorithm takes a similar time as our approach for the same sampling radius (both methods take about $0.3$s for $r\approx\,0.012$). However, as shown in Figure~\ref{fig:cmp}(a), we are able to detect the gaps from their output and show that this competing result is not maximal. In another comparison of uniform sampling, the Elk model is remeshed with $31k$ vertices. We set the desired angle bound to $[37^o,98^o]$ and still obtain a higher quality output, compared with the state of the art~\cite{Yan2009}. The results are summarized in Table~\ref{tab:remesh}.

\begin{figure*}[!htp]
\centerline{
\includegraphics[width=1.0\linewidth]{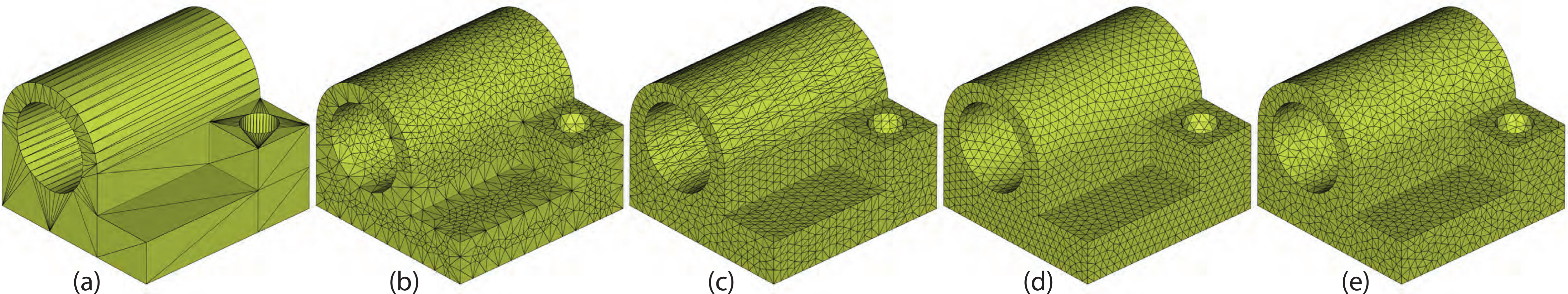}
}
\caption{Comparison of the remeshed Joint model, 3.2k vertices. (a) Input mesh, (b) DelPSC~\protect\cite{Cheng07b}; (c) CVD~\protect\cite{Valette2008}; (d) CVT~\protect\cite{Yan2009}; (e) ours (the model is from the Aim@Shape repository).}
\label{fig:joint}
\end{figure*}

We compare our feature-preserving (uniform) remeshing result with previous work in Figure~\ref{fig:joint}. In this example, we request a minimal angle of $35^o$, but we can only obtain $33.9^o$. We also cannot get a pure valence $5,6,7$ solution because of the feature constraints.

\paragraph*{Adaptive remeshing}

\begin{figure*}[t]
\centerline{
\includegraphics[width=1.0\textwidth]{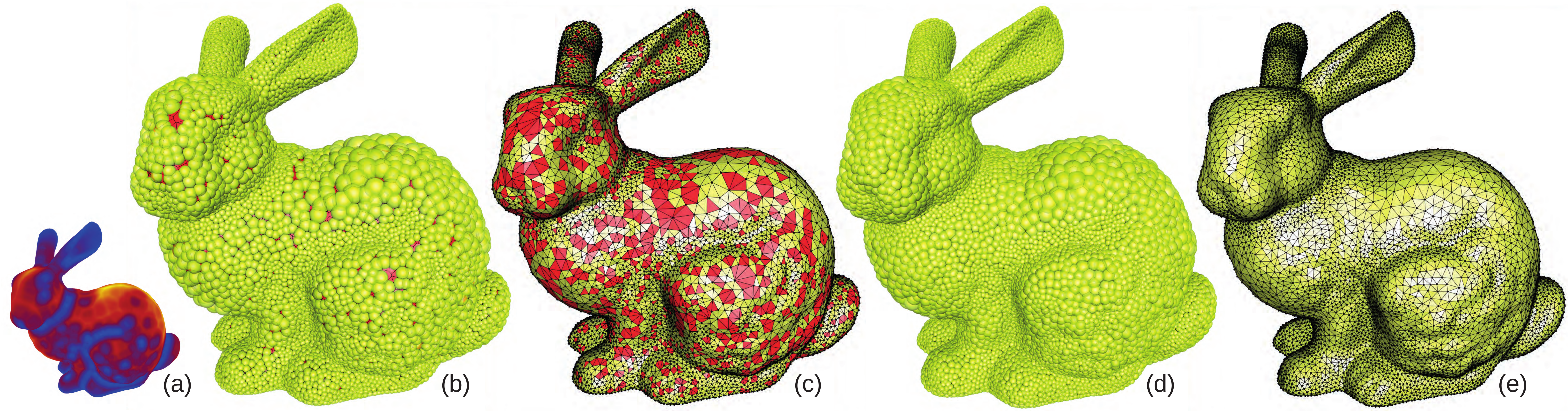}
}
\caption{Adaptive maximal Poisson-disk sampling on the Bunny model. (a) the density map, where cooler colors correspond to a smaller radius while warmer colors correspond to a larger radius; (b) non-maximal sampling results in gaps (red regions); (c) triangles of the remeshing that are effected by gaps (red triangles); (d) maximal sampling; and (e) remeshing using maximal sampling. Only maximal sampling leads to good triangle shapes that are essential for simulation (the model is from  the Stanford 3D scanning repository).}
\label{fig:teaser}
\end{figure*}

\begin{figure}[!htp]
\centerline{\includegraphics[width=1.0\linewidth]{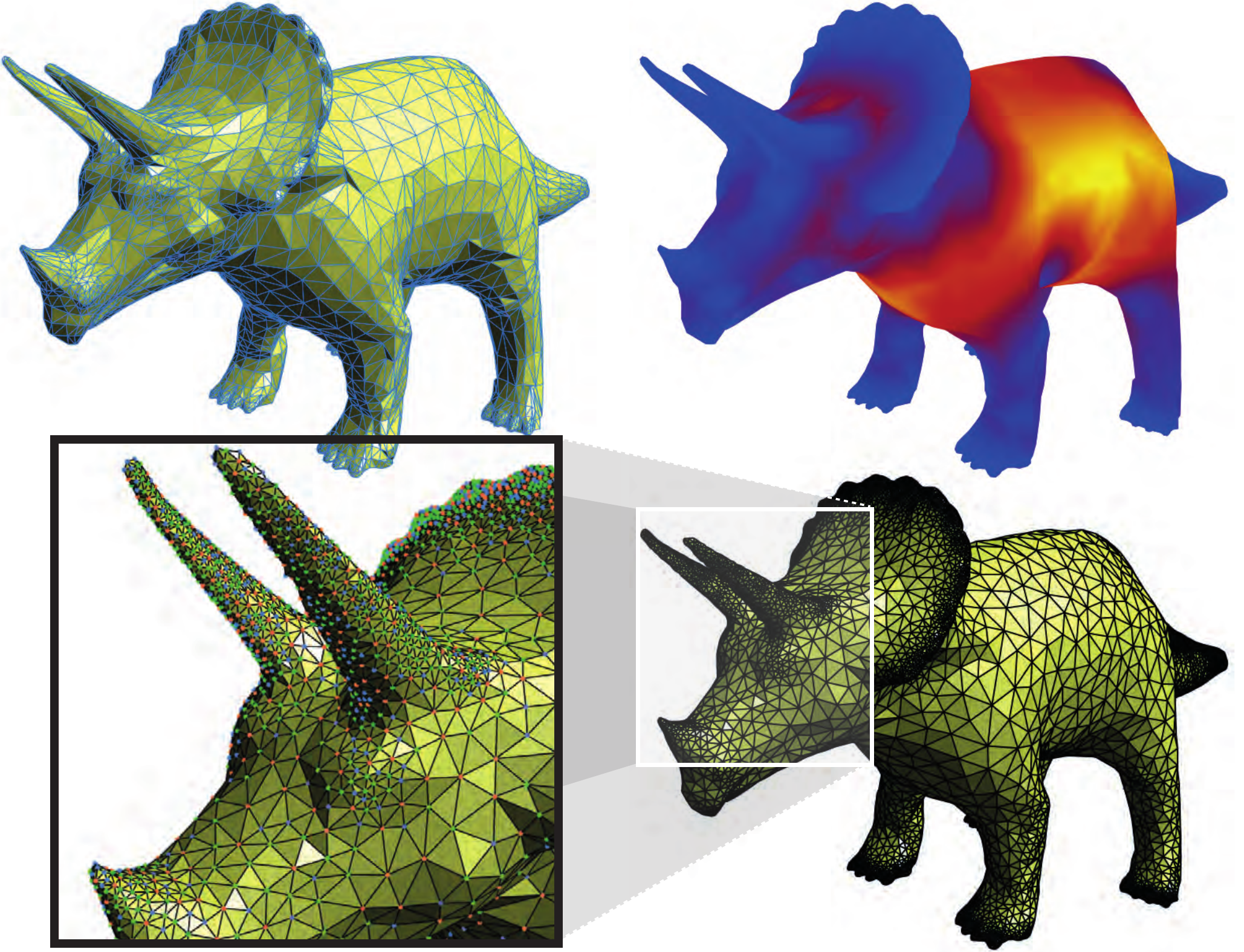}}
\caption{Adaptive sampling/remeshing. Top: input mesh and the density function; Bottom: remeshing result (the model is from the Aim@Shape repository).}
\label{fig:remesh}
\end{figure}

\begin{figure}[!htp]
\centerline{
\includegraphics[width=1.0\linewidth]{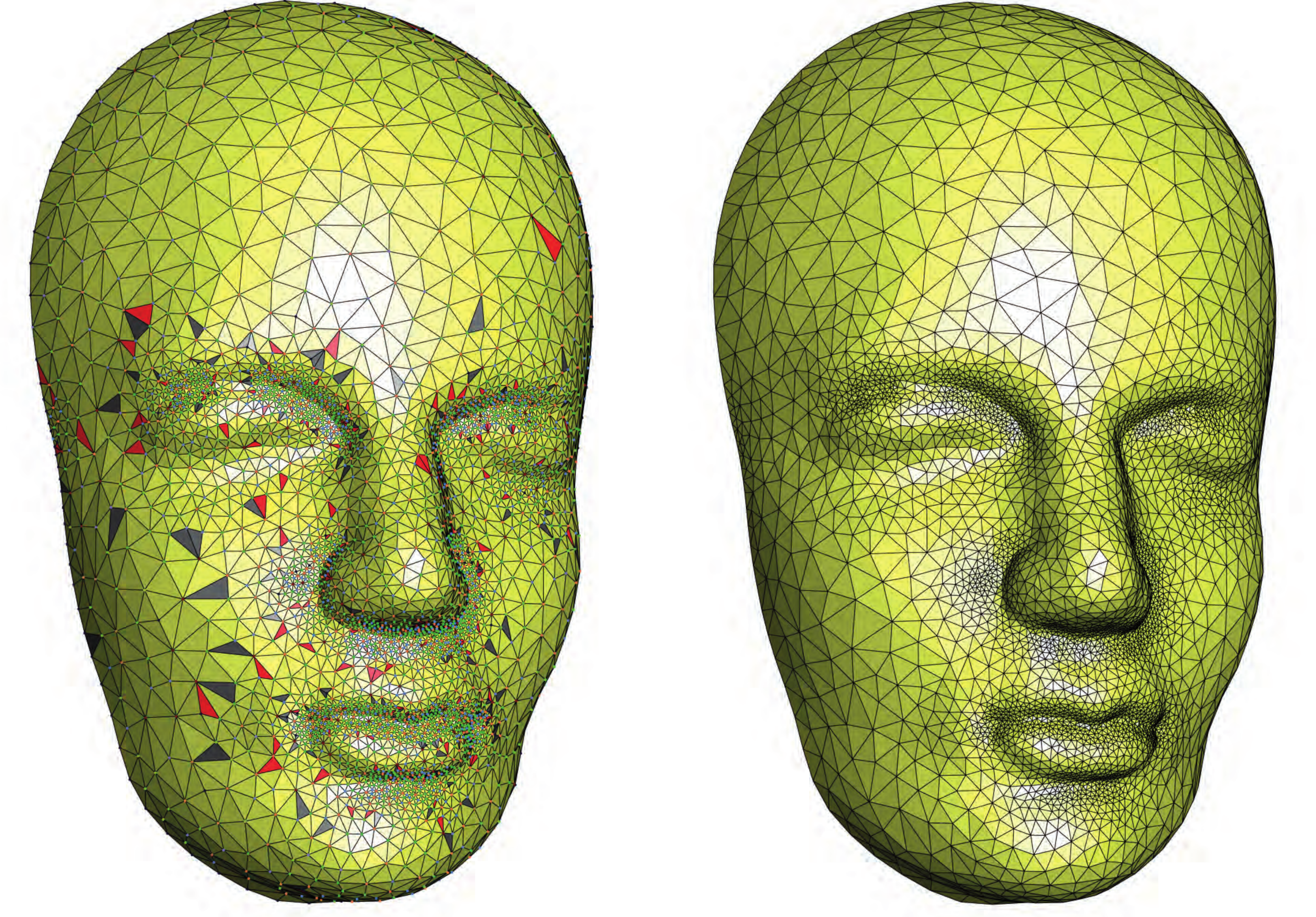}
}
\caption{Boundary handling, 9k vertices. (a) maximal sampling/remeshing. The triangles with $\theta_{min}<30^o$ are shown in dark gray and the triangles with $\theta_{min}\in[30^o,32^o]$ are shown in red. (b) Optimized sampling and remeshing. The optimized angle range is $[32^o,115^o]$ (the model is from the Aim@Shape repository).}
\label{fig:mask}
\end{figure}

We applied our adaptive remeshing/optimization to various models. We use the \emph{local feature size} (lfs)~\cite{Amenta1998} as the density function, i.e., $\rho(\mx)=\frac{1}{{lfs}(\mx)^2}$. Several results are shown in Figure~\ref{fig:remesh} and ~\ref{fig:teaser}. Another example of adaptive sampling on a mesh surface with boundaries is shown in Figure~\ref{fig:mask}. The boundary is treated the same as a 1D feature curve.

We compare our remeshing algorithm with the state-of-the-art remeshing approaches~\cite{Cheng07b,Valette2008,Fu2008,Yan2009}, in terms of $Q_{min}$, $\theta_{min}$, $\theta_{max}$, Hausdorff and RMS distances, the ratio of angles smaller than $30^o$ and the ratio of the valence 5,6,7 vertices (see Table~\ref{tab:remesh}). In Table~\ref{tab:remesh}, SAG refers to \protect\cite{Surazhsky03}, DEL refers to \protect\cite{Cheng07b}, DSR refers to \protect\cite{Fu2008}, CVD refers to \protect\cite{Valette2008}, CVT refers to \protect\cite{Yan2009}, CAP refers to \protect\cite{Chen2012}, and MPS refers to our approach without optimization. In adaptive remeshing, the desired angle bound is set to $[32^o, 115^o]$. Similar to the uniform meshing case, we can observe that our approach exhibits better $Q_{min}$ and $\theta_{min}$, as well as high approximation quality to the input surface. For CVT and CAP, we use 100 iterations to generate all the results shown in this paper. Our method has a larger Hausdorff distance compared with CVT-based remeshing~\cite{Yan2009} due to the fact that we did not explicitly optimize the Hausdorff distance, while CVT tends to approximate the input mesh by minimizing the Hausdorff distance~\cite{Nivoliers2011}. We would like to address this issue in future work using our randomized optimization framework.

\begin{table*}[!htp]
\centering
{
\begin{tabular*}{0.9\linewidth}{@{\extracolsep{\fill}}|c|c|c|c|c|c|c|c|c|c|}
\hline Model &  \#v   &$Q_{min}$& $\theta_{min}$ & $\theta_{max}$  & Hdist & RMS & $\theta_{<30^o}$&$v_{567}$&Time(s)\\
\hline \multicolumn{9}{|c|}{adaptive remeshing}& \\
\hline Rockerarm[CVD] & 5.8k & 0.353 & 16.5   & 133.2 & 0.44 & 0.048 & 0.30 & 98.1 & 3.0\\
\hline Rockerarm[DSR] & 5.8k & 0.004 & 0.14   & 152.9 & 0.55 & 0.057 & 1.34 & 93.1 & 35.0 \\
\hline Rockerarm[CVT] & 5.8k & 0.428 & 21.3   & 124.4 & \bf{0.34} & \bf{0.032} & 0.11 & 99.7 & 47.5 \\
\hline Rockerarm[CAP] & 5.8k & 0.413 & 23.5   & 126.4 & 0.58 & 0.050 & 0.12 & 98.8 & 59.8\\
\hline Rockerarm[MPS] & 5.3k & 0.320 & 14.0   & 130.0 & 0.48 & 0.034 & 1.49 & 90.6 & 3.2\\
\hline Rockerarm[OUR] & 5.8k & \bf{0.516} & \bf{32.0}   & \bf{113.6} & 0.48 & 0.033 & \bf{0} & \bf{100} & 4.8 \\
\hline
\hline Homer[CVD]  & 7.2k & 0.049 & 2.82   & 173.2 & 0.35  & 0.057 & 2.03 & 94.3 & 3.1\\
\hline Homer[DSR]  & 7.5k & 0.150 & 21.7   & 129.9 & 0.43  & 0.068 & 0.10 & 95.2 & 67.0\\
\hline Homer[CVT]  & 7.2k & 0.334 & 16.2   & 120.8 & \bf{0.20}  & \bf{0.029} & 2.04 & 99.5 & 74.2\\
\hline Homer[CAP]  & 7.2k & 0.405 & 21.0   & 126.6 & 0.23  & 0.046 & 0.42 & 96.4 & 81.9 \\
\hline Homer[MPS]  & 7.3k & 0.371 & 21.8   & 131.4 & 0.32  & 0.028 & 0.29 & 95.4 & 3.5\\
\hline Homer[OUR]  & 7.2k & \bf{0.513} & \bf{32.0} & \bf{115.0} & 0.31  & \bf{0.023} & \bf{0}  & \bf{100} & 5.1 \\
\hline
\hline Triceratops[CVD]  & 9k   & 0.007 & 0.46   & 179.1 & 0.59 & 0.050 & 4.37  & 94.1 & 4.8\\
\hline Triceratops[CVT]  & 9k   & 0.385 & 15.5   & 127.3 & \bf{0.24} & \bf{0.038} & 0.29  & 99.2 & 82.9\\
\hline Triceratops[CAP]  & 9k   & 0.411 & 21.2   & 126.4 & 0.32 & 0.035 & 0.43  & 97.2 & 116.4\\
\hline Triceratops[MPS]  & 9k   & 0.29 & 13.5   & 138.6 & 0.48 & 0.062 & 1.23  & 92.0  & 6.3 \\
\hline Triceratops[OUR]  & 9k   & \bf{0.506} & \bf{32.0}   & \bf{114.8} & 0.46 & 0.062 & \bf{0} & \bf{100}  & 28.4\\
\hline
\hline Elephant[CVD] & 11k & 0.040 & 1.33  & 173.2 & 0.38 & 0.039 & 4.15 & 95.4 & 7.2 \\
\hline Elephant[CVT] & 11k & 0.408 & 21.2  & 126.6 & \bf{0.23} & \bf{0.024} & 0.72 & 94.0 & 102.5 \\
\hline Elephant[CAP] & 11k & 0.316 & 18.5  & 138.1 & 0.24 & \bf{0.024} & 0.04 & 93.7 & 115.6 \\
\hline Elephant[MPS] & 11k & 0.301 & 13.0  & 130.0 & 0.46 & 0.029 & 1.09 & 92.6 & 8.4 \\
\hline Elephant[OUR] & 11k & \bf{0.505} & \bf{32.0}   & \bf{114.9} & 0.38 & 0.061 & \bf{0}    & \bf{100}  & 30.6\\
\hline
\hline Bunny[CVD]  & 12k   & 0.15 & 9.6   & 160.1 & 0.34 & \bf{0.028} & 0.71  & 96.0 & 5.2\\
\hline Bunny[CVT]  & 12k   & 0.36 & 17.8  & 133.2 & \bf{0.20} & 0.029 & 0.38  & 96.4 & 176.3\\
\hline Bunny[CAP]  & 12k   & 0.20 & 7.48   & 137.8 & 0.48 & 0.038 & 4.99  & 97.5 & 185.0\\
\hline Bunny[MPS]  & 9.8k  & 0.36 & 19.2 & 133.2 & 0.46 & 0.042 & 0.90 & 94.7 & 18.8\\
\hline Bunny[OUR]  & 12k   & \bf{0.51} & \bf{32.0}  & \bf{114.6} & 0.37 & 0.035 & \bf{0}  & \bf{100} & 23.7\\
\hline \multicolumn{9}{|c|}{uniform remeshing} & \\
\hline Joint[CVD] & 3.2k & 0.040 & 2.4    & 174.6 & \bf{0.12}  & \bf{0.011} & 15.6 & 96.0 & 2.5\\
\hline Joint[DEL] & 3.2k & 0.057 & 2.83   & 171.5 & 0.38  & 0.031 & 2.6  & 91.1 & 8.9\\
\hline Joint[CVT] & 3.2k & 0.555 & 29.6   & 108.4 & 0.26  & 0.056 & 0.005& 99.2 & 23.7 \\
\hline Joint[OUR] & 3.2k & \bf{0.688} & \bf{33.9}   & \bf{104.9}  & 0.37  & 0.056 & \bf{0} & \bf{99.4} & 4.5\\
\hline
\hline Bunny[CVD] & 12k   & 0.111 & 3.81   & 146.2 & 0.43 & 0.037 & 17.2  & 97.6 & 4.3 \\
\hline Bunny[CVT] & 12k   & 0.618 & 37.4   & 101.2 & \bf{0.20} & \bf{0.029} & \bf{0}  & \bf{100} & 171.9\\
\hline Bunny[CAP] & 12k   & 0.215 & 20.5   & 126.0 & 0.35 & \bf{0.029} & 0.41  & 99.5 & 181.2\\
\hline Bunny[MPS] & 12k   & 0.480 & 30.3   & 118.0 & 0.44 & 0.037 & 0  & 96.5 & 10.6\\
\hline Bunny[OUR] & 12k   & \bf{0.629} & \bf{38.0}   & \bf{100.0} & 0.47 & 0.034 & \bf{0}  & \bf{100} & 56.8\\
\hline
\hline Elk[SAG] & 31k & 0.092 & 4.72   & 166.8 & 0.30  & 0.047 & 0.07 & 99.7 & 60.0 \\
\hline Elk[CVD] & 31k & 0.037 & 2.28   & 175.0 & \bf{0.20}  & \bf{0.012} & 1.60 & 96.5 & 6.1\\
\hline Elk[CVT] & 31k & 0.634 & 36.5   & 99.2  & 0.21  & 0.014 & \bf{0}  & 99.9 & 215.3\\
\hline Elk[CAP] & 31k & 0.305 & 11.4   & 133.4  & 0.30  & 0.027 & 1.93   & 98.7 & 242.8\\
\hline Elk[MPS] & 31k & 0.475 & 30.1   & 118.6  & 0.22  & 0.020 & \bf{0}    & 96.4 & 6.3 \\
\hline Elk[OUR] & 31k & \bf{0.645} & \bf{37.0}   &\bf{97.9}  & 0.26  & 0.020 & \bf{0} & \bf{100} & 87.5 \\
\hline
\end{tabular*}
} \caption{Statistics of the remeshing quality compared with previous work. $\#v$ is the number of sampled points. $Q_{min}$ is the minimal triangle quality~\protect\cite{Frey1997}; $\theta_{min}$ and $\theta_{max}$ are the minimal and maximal angle; Hdist and RMS are the Hausdorff distance and the root mean square distance between remeshing and the input mesh (\% of the diagonal length of the bounding box); $v_{567}$ is the percent of vertices with valences 5,6 and 7. The best result is highlighted with bold font.}
\label{tab:remesh}
\end{table*}

\begin{figure}[ht]
\centerline{
\includegraphics[width=0.49\linewidth]{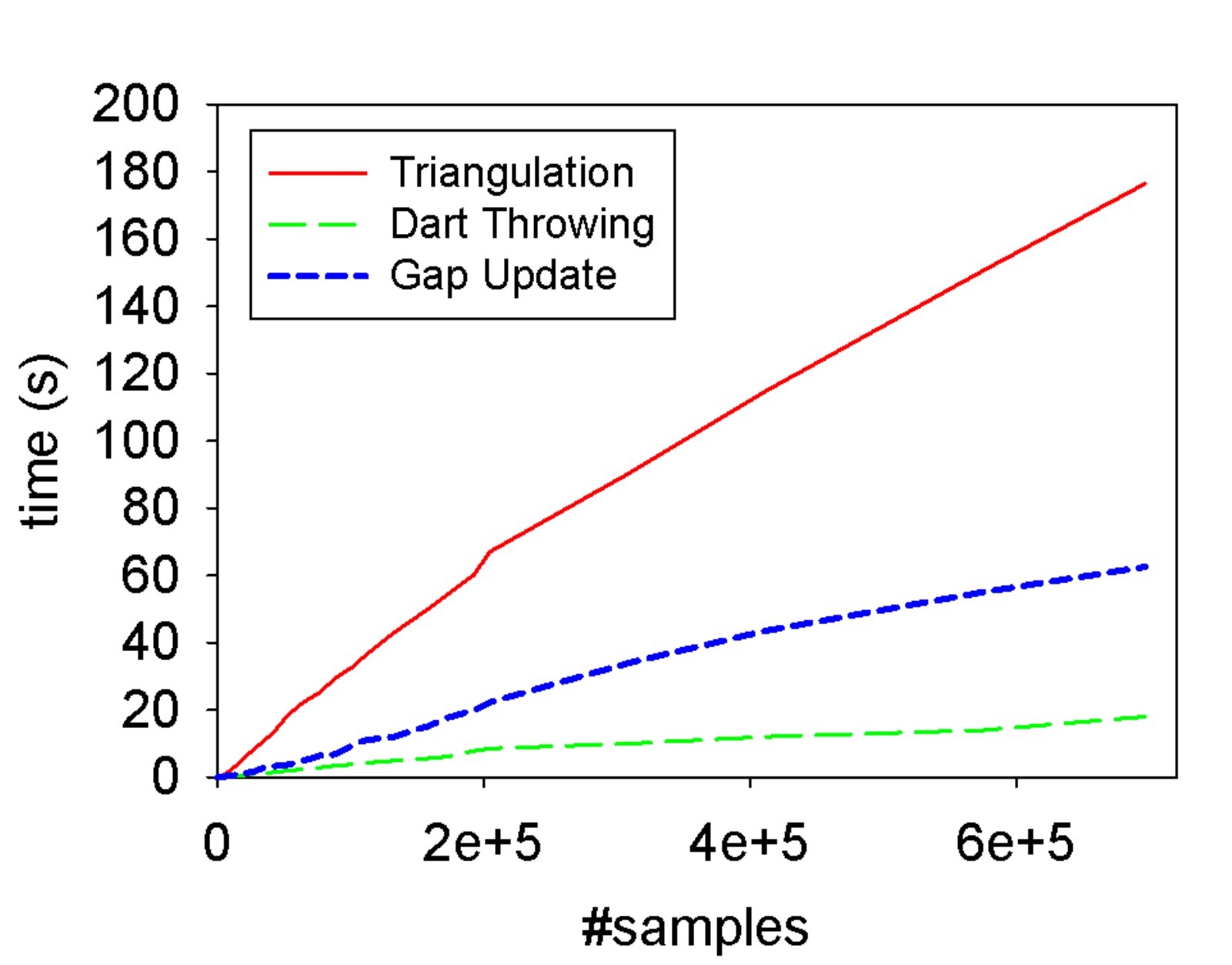}
\includegraphics[width=0.51\linewidth]{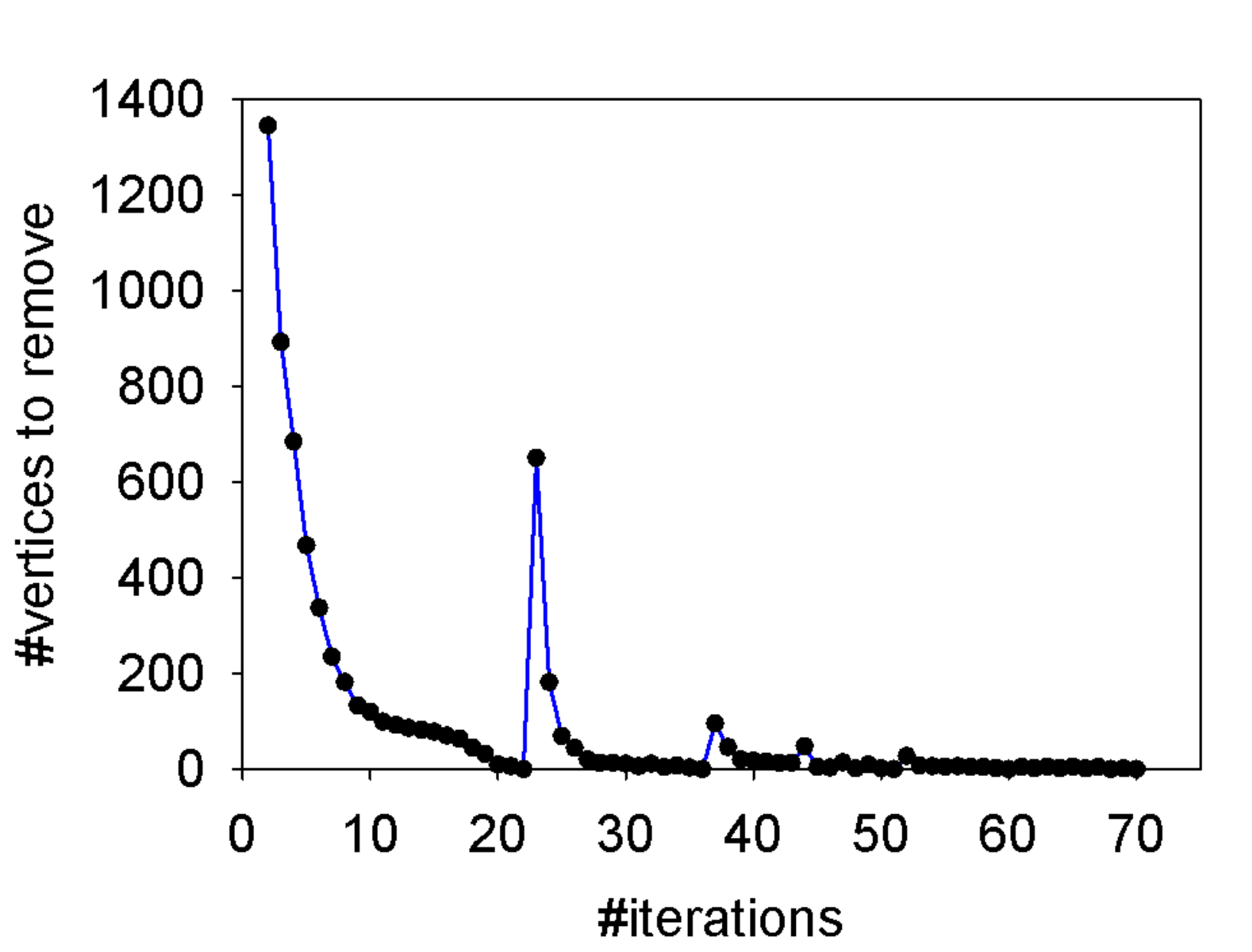}
}
\caption{Left: Timing curves of our sampling algorithm on the Bunny model. Right: convergence curve of the valence/angle optimization of Figure~\protect\ref{fig:teaser}. Each peak of the curve corresponds to a switch between valence/angle optimization. }
\label{fig:timing_surf}
\end{figure}

\paragraph*{Efficiency}
Figure~\ref{fig:timing_surf}(a) shows the timing curve of our algorithm with an increasing number of sample points, and the convergence of the valence/angle optimization is shown in Figure~\ref{fig:timing_surf}(b) on the Bunny model shown in Figure~\ref{fig:teaser}. For this example, our algorithm takes $12.4s$ for the initial sampling, $6.4s$ for gap filling and $4.9s$ for optimization ($23.7s$ in total), while CVT takes $182s$ and CAP takes $391s$ for the same number of samples.

\paragraph*{Limitations}
There are several limitations to our current approach that can be addressed in future work. 1) Although the sampling framework works well for mesh surfaces, we might fail to compute a valid remeshing if the minimal radius is larger than the local feature size of the surface. For example, we can remesh a thin sheet mesh with parallel planes using small triangles, but we cannot remesh it using large triangles.
Another limitation is that we do not have any theoretical guarantee for convergence of the randomized mesh optimization. This is partially due to the fact that in some extreme cases the valence and angle optimization conflict with each other. One possible solution is to enlarge the region further to be resampled. The third limitation is that our framework depends on the regular triangulation/power diagram, which limits the performance of our algorithm compared with GPU-based approaches~\cite{Bowers2010}.

\section{Conclusion and Future Work} \label{sec:con}

We presented a theoretical analysis of gaps in disk sets with varying radii in arbitrary dimensions. Based on this analysis, we proposed algorithms and data structures for gap detection and gap updates when the disk set changes in the 2D plane and on 3D surfaces. The contribution of our work is illustrated with an adaptive remeshing algorithm that can improve the remeshing quality in aspects of minimal angle, vertex valence and triangle quality. In future work, we would like to build a solution for volumetric remeshing that can be applied to static and deformable objects and to extend our gap processing framework to higher dimensions.

\begin{acks}
We would like to thank the anonymous reviewers for their detailed comments. We thank Helmut Pottmann, Xiaohong Jia and Rachid Ait-Haddou for their valuable discussions, Virginia Unkefer and Sawsan Al-Halawani for proof-reading and all the other GMSV members for their enthusiastic help during this work. We are grateful to Thouis R. Jones, Manuel N. Gamito and David Cline for sharing their code, Tamal K. Dey, S\'ebastien Vallete and Zhonggui Chen for their software, and Pierre Alliez and Bingfeng Zhou for their results.
\end{acks}

\bibliographystyle{acmtog}
\bibliography{AMPS}

\appendix

\section{Correctness of gap decomposition}
\label{app:decom}

In this appendix, we prove that the primitive extraction algorithm presented in Section~\ref{sec:extract} gives a valid decomposition of a gap.

\begin{figure*}[!ht]
\centerline{
\hfill\includegraphics[width=.32\linewidth]{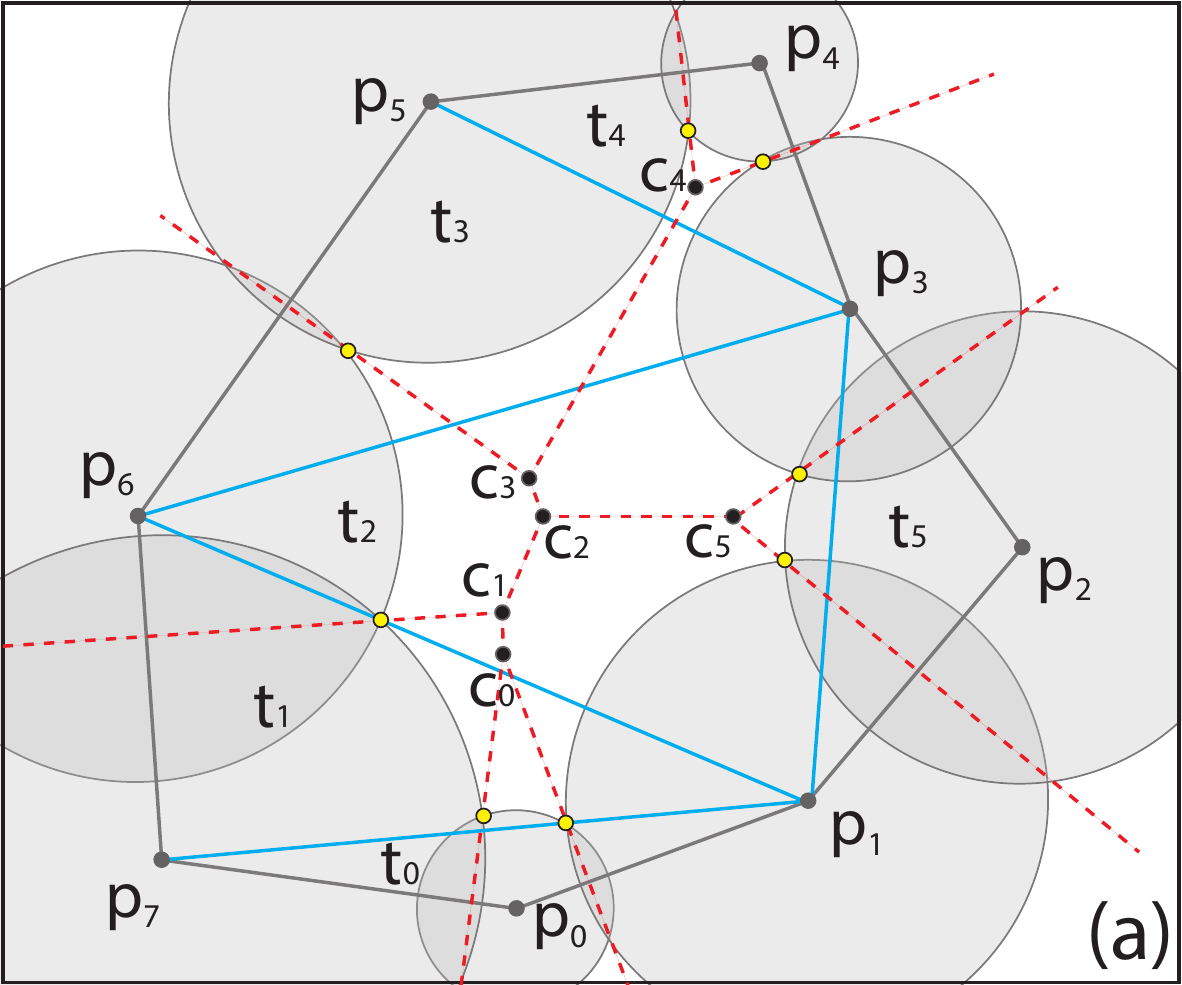}
\hfill\includegraphics[width=.32\linewidth]{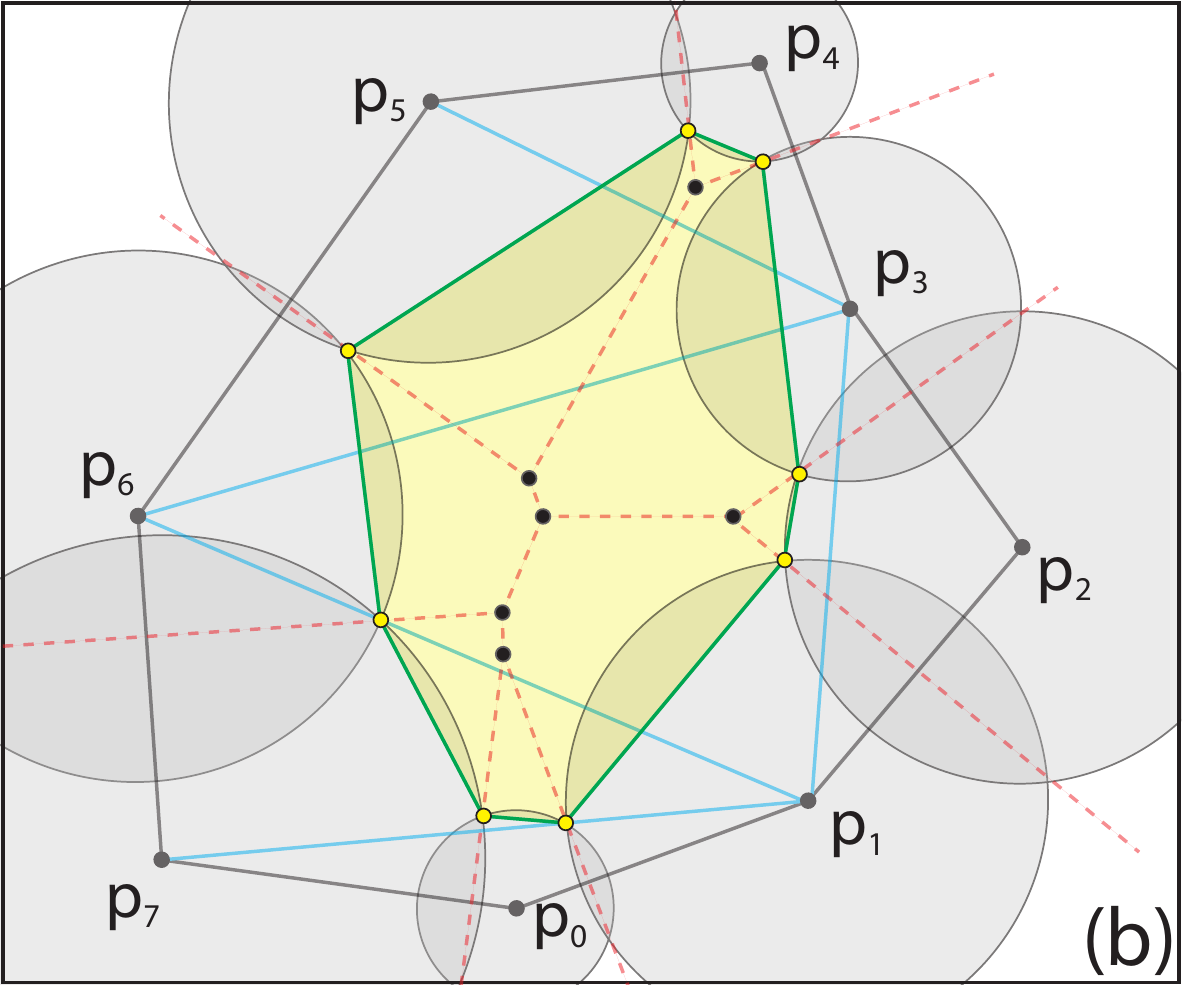}
\hfill\includegraphics[width=.32\linewidth]{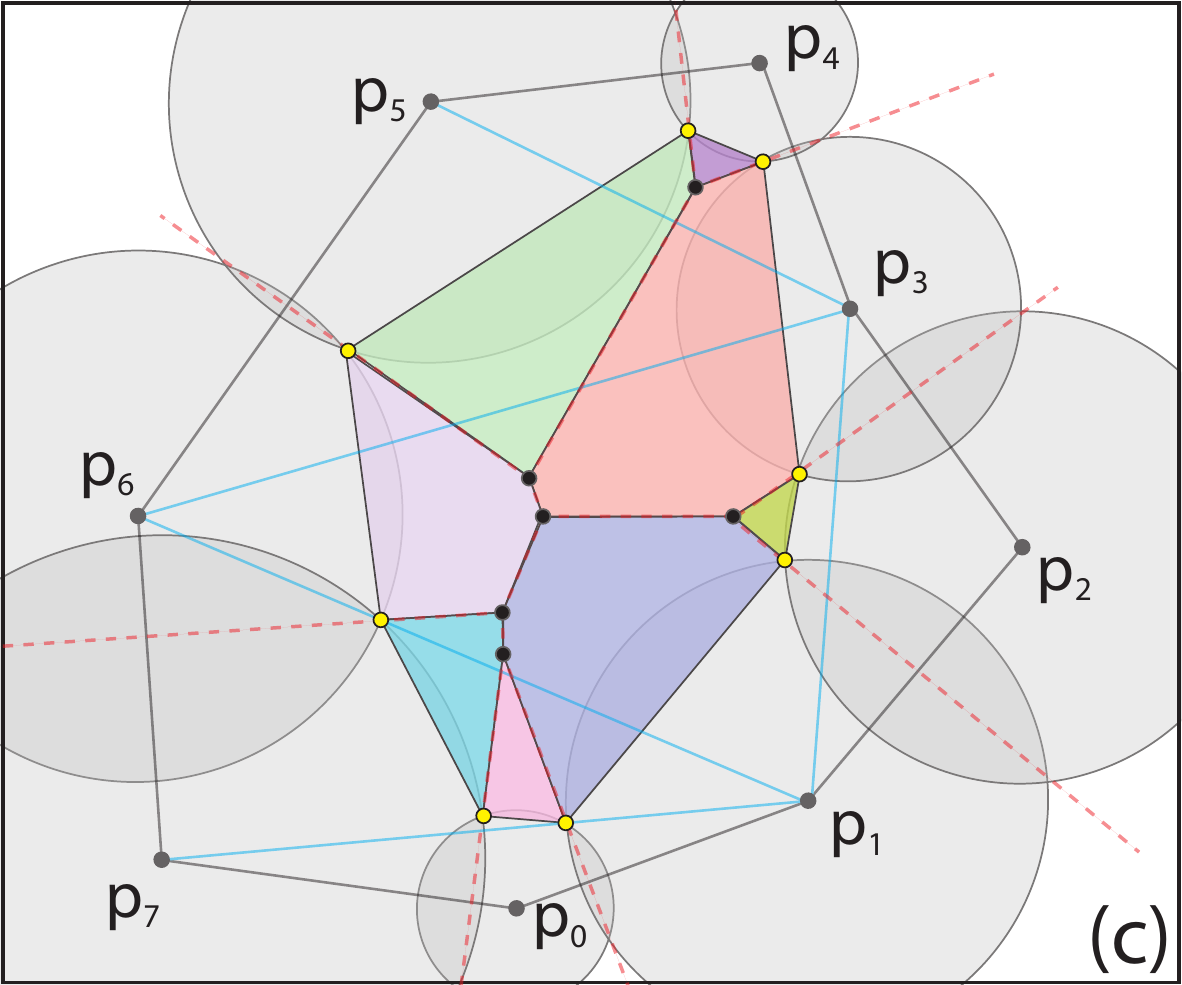}
\hfill}
\centerline{
\hfill\includegraphics[width=.32\linewidth]{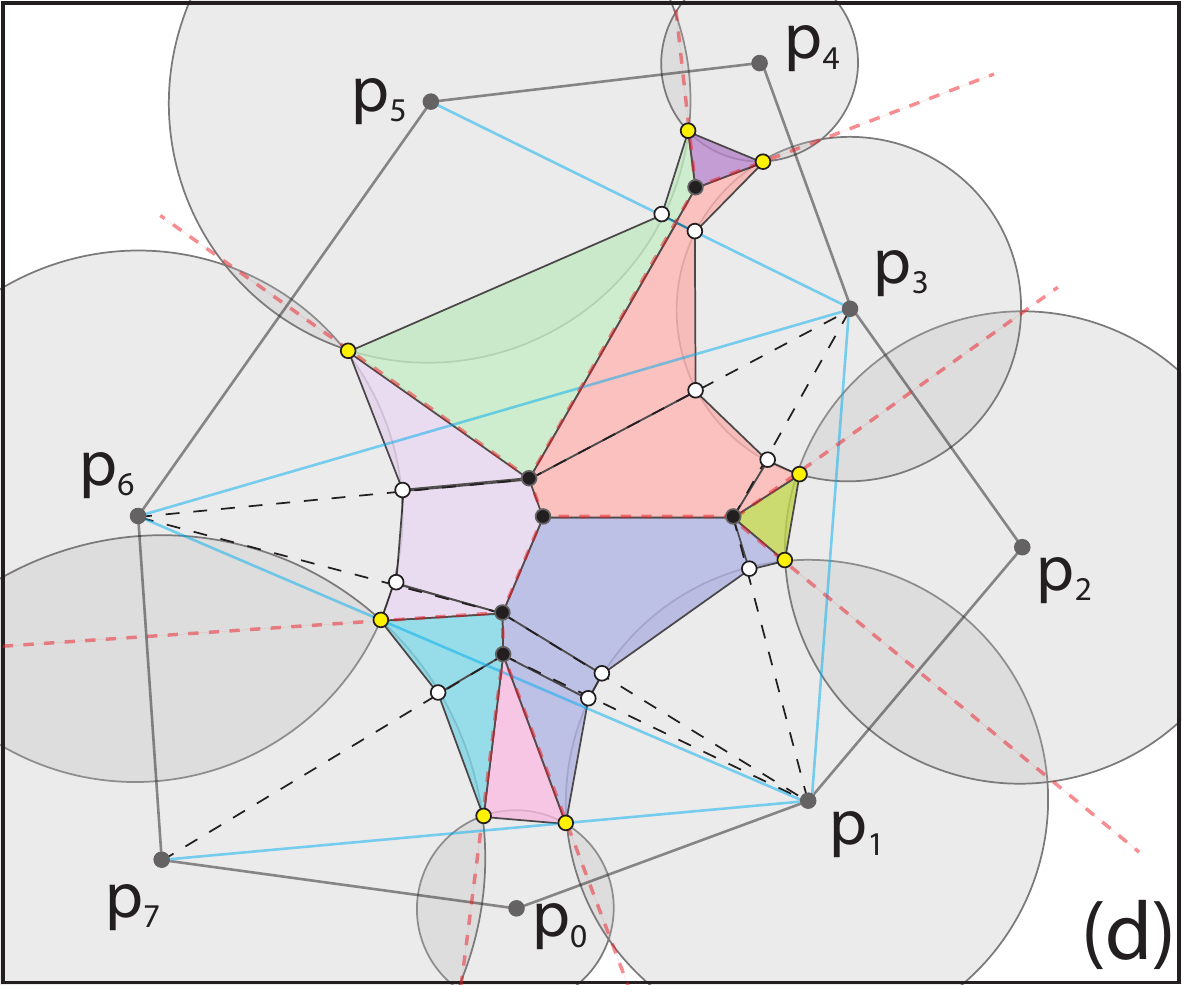}
\hfill\includegraphics[width=.32\linewidth]{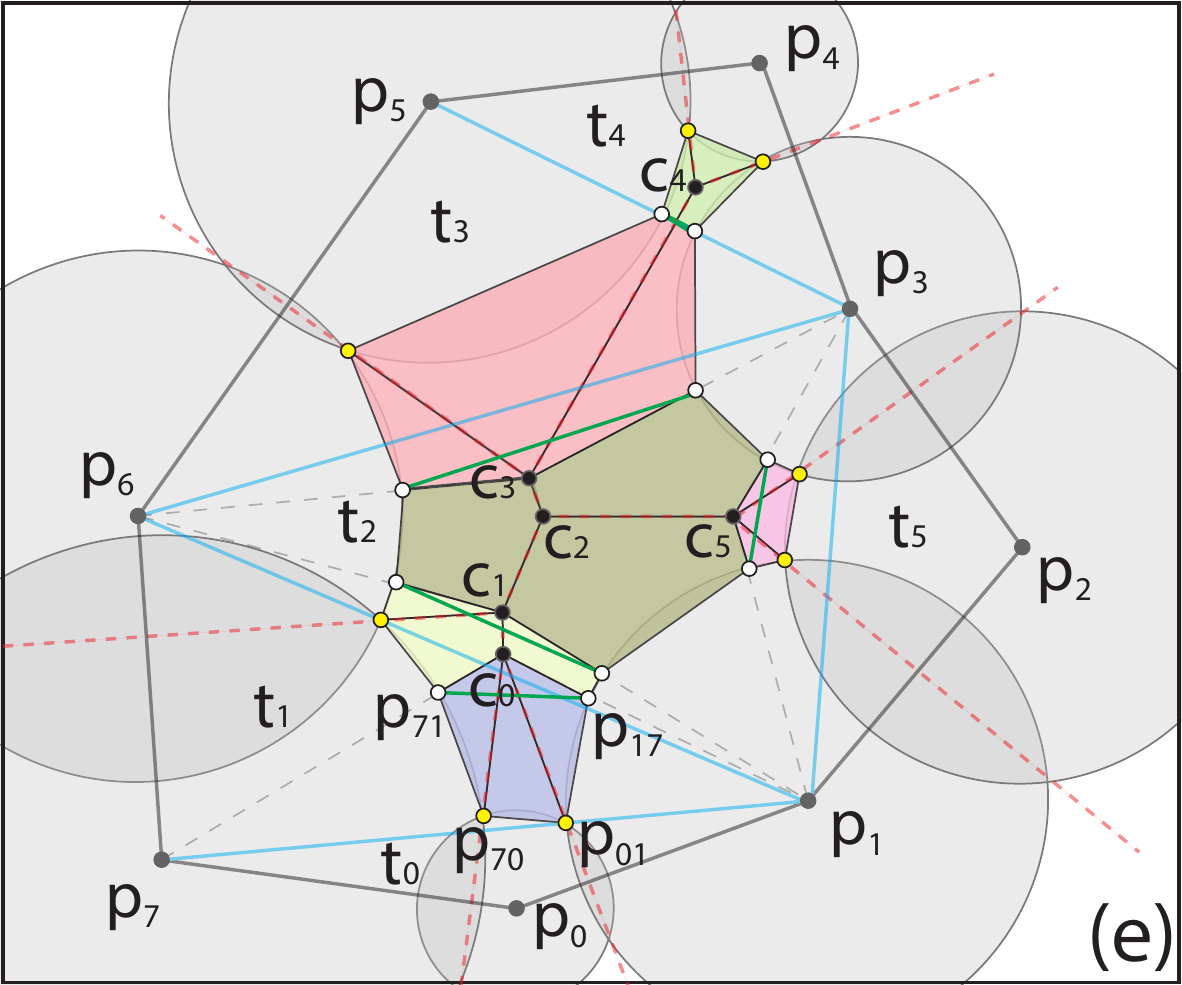}
\hfill\includegraphics[width=.32\linewidth]{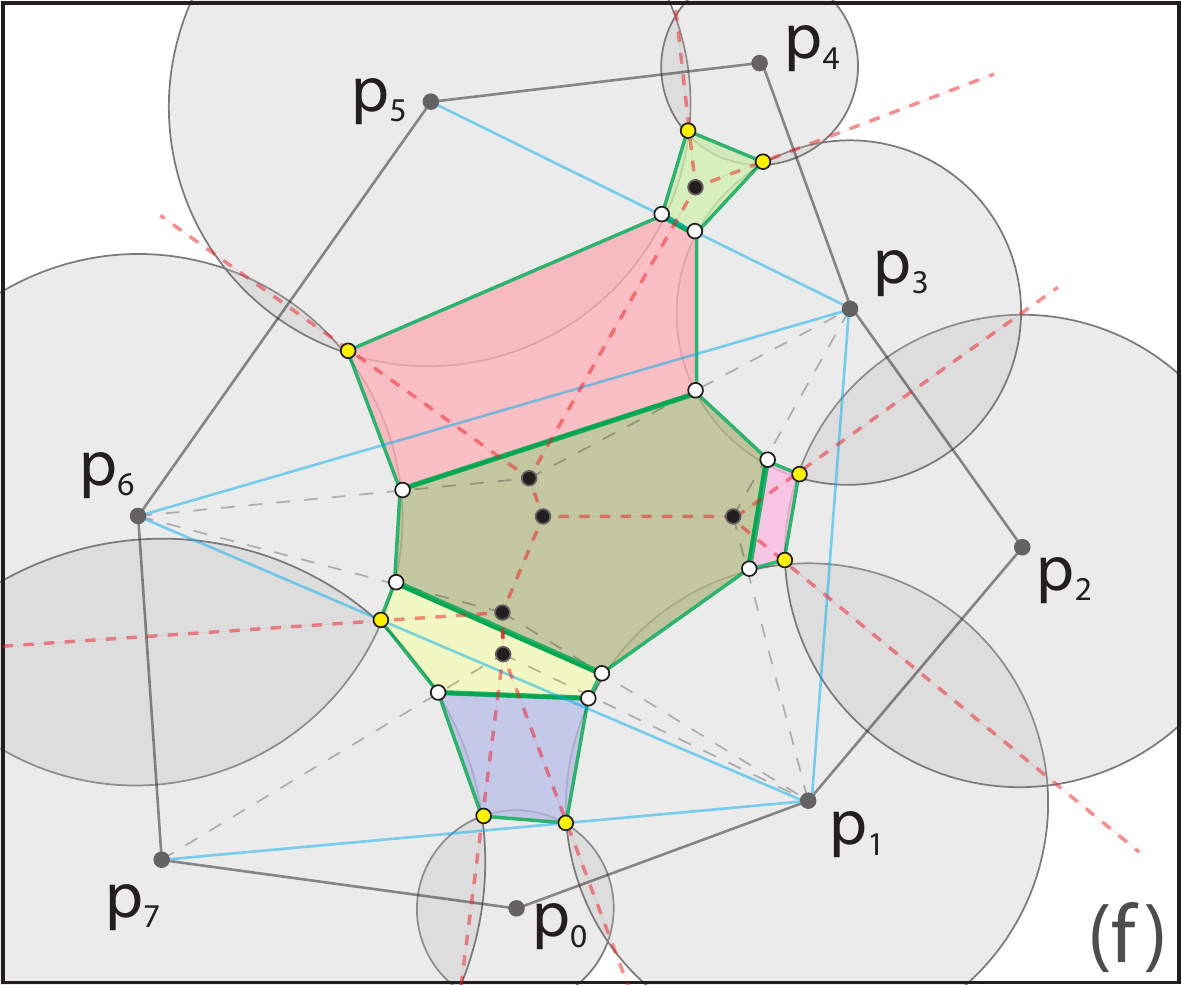}
\hfill}
\caption{An illustration our gap decomposition algorithm. (a): power diagram (red) and regular triangulation (solid black and blue) of a connected gap component. Five power centers fall into the same triangle $t_2$; (b) initial gap polygon; (c) the gap polygon is subdivided into several gap cells; (d) each gap cell is split into deg$(\mp_i)$ sub-cells; (e) the sub-cells are regrouped to from tri-cells for each triangle; (f) the tri-cells are split and merged by auxiliary splitting edges.}
\label{fig:appendix}
\end{figure*}

\subsection{Properties of gaps}

Recall that a connected gap component is incident to a set of gap triangles. The gap triangles of a gap can be clustered together using the first three connectivity rules (Section~\ref{sec:cluster}). In the following discussion, when we refer to a triangle, we mean that the triangle is a gap triangle that is incident to the same gap.

The edges of gap triangles can be classified into two types, boundary edges (solid black) and inner edges (blue), as illustrated in Figure~\ref{fig:appendix}(a). A boundary edge has an edge length smaller than the sum of the radii of its two endpoints, i.e., $|\mp_i\,\mp_j|<r_i+r_j$, which means that the edge is covered by the two disks. The boundary edges isolate the gap from other gaps. Each inner edge is shared by two neighboring gap triangles. The inner edges can be further classified into two types: (a) inner edges with two power centers of neighboring gap triangles that lie on different sides of the edge, e.g., edge $\mp_3\,\mp_5$ in Figure~\ref{fig:appendix}(b); (b) inner edges with two power centers of neighboring gap triangles that lie on the same sides of the edge, e.g., edges $\mp_1\,\mp_7$, $\mp_1\,\mp_6$,$\mp_1\,\mp_3$ and $\mp_3\,\mp_6$.

A gap is bounded by a set of circular arcs. The vertices of the arcs are the intersection points of the disks (yellow points in Figure~\ref{fig:appendix}). Each vertex of the arcs is incident to a boundary edge of the regular triangulation. Note that no arc intersects with other arcs except with two neighboring arcs at the arc vertices (otherwise the gap is not connected). The arcs are simply connected. Without loss of generality, we assume the ccw orientation, such that the gap is on the left side when we traverse the arcs, see Figure~\ref{fig:appendix}(a) for an example.

If we replace each arc with a segment, then the resulting polygon covers the gap. The polygon is called a
\emph{gap polygon}, as shown in Figure~\ref{fig:appendix}(b).  As discussed in Section~\ref{sec:overview}, the gap is defined as the difference between the domain and the disks, i.e., $\Omega-\cup_i\{(\mp_i,r_i)\} = \cup\{\Omega_i\}-\cup_i\{(\mp_i,r_i)\}$, which is equivalent to the union of each power cell minus its incident disk, i.e., $\cup_i\{\Omega_i-(\mp_i,r_i)\}$. Each power cell contains exactly one arc of the gap boundary. The gap polygon can be decomposed by the power cells (intersecting the gap polygon with the power cell of each disk), as shown in Figure~\ref{fig:appendix}(c). We call each such (colored) polygon a \emph{gap cell} of this disk.

\subsection{Proof of the validness}

We prove the correctness of our algorithm by constructing a valid decomposition of the gap polygon and show that our algorithm is equivalent to this valid decomposition. We define deg$(\mp_i)$ as the number of incident triangles of each disk. Hence, each disk is incident to two boundary edges and deg$(\mp_i)-1$ inner edges. We first split the gap cell of each disk into deg$(\mp_i)$ sub-cells by inserting deg$(\mp_i)-1$ auxiliary vertices on each arc, such that each sub-cell is associated with an incident triangle of the disk. For each type (a) inner edge of a disk, we simply split the arc by the edge, and we also split the power edge by the edge, e.g., edge $\mp_5\mp_3$; For each type (b) inner edge of a disk, we split the arc by connecting the disk center and the power vertex incident to the edge.

After this splitting process, each gap cell is split into deg$(\mp_i)$ sub-cells, each sub-cell is a convex polygon and the union all the sub-cells covers the gap, as shown in Figure~\ref{fig:appendix}(d). Until now, all the sub-cells are associated with disks. Next, we reassign the sub-cells to triangles. Since each gap cell is split into deg$(\mp_i)$ sub-cells, we associate each sub-cell to an incident triangle of the disk. After the regrouping, each triangle has three sub-cells, and each pair of sub-cells shares a common power edge, as shown in Figure~\ref{fig:appendix}(e).
We call the union of the three sub-cells associated with a triangle a \emph{tri-cell}. The tri-cell of a triangle has the following properties:
\begin{itemize}
\item Each tri-cell is a simple polygon in ccw orientation.
\item The vertices on the arcs of a tri-cell are convex in ccw orientation (white and yellow vertices).
\item The tri-cell is convex if the triangle has one or no type (b) inner edge.
\item A power center of a tri-cell contributes to a convex vertex to its own triangle and contributes to a concave vertex to its neighboring triangle (Figure~\ref{fig:appendix}(e)).
\end{itemize}

There are two types of vertices on the arc, the intersection point of two neighboring arcs (yellow) and the auxiliary vertices (white) by intersecting the disk center to its power vertices. An auxiliary vertex is convex since it is from a convex sub-cell by our splitting process. To see that the yellow vertices are convex, we use the tri-cell of $t_0$ as an illustration, as shown in Figure~\ref{fig:appendix}(e). There are two arc-arc vertices $\mp_{70}$ and $\mp_{01}$ and two auxiliary vertices $\mp_{71}$
and $\mp_{17}$. Let us look at the quad $\mc_0\mp_7\mp_0\mp_7$, since $\mc_0$ is the power center of triangle $\mp_7\mp_0\mp_1$, which lies outside of the triangle. Hence, $\mc_0$ must lie on the left side of edge $\mp_{70}\mp_{01}$. Since $\mp_{17}$ lies on the left side of the power edge $\mc_{0}\mp_{01}$, and similarly $\mp_{71}$ lies on the right side of the power edge $\mc_{0}\mp_{70}$, and $\mc_0$ lies on the right side of $\mp_{17}\mp_{71}$, we can see that the tri-cell $\mc_{0}\mp_{71}\mp_{70}\mp_{01}\mp_{17}$ is a convex polygon. Hence the arc-arc vertices are convex. The convexity of the other arc-arc vertices can be explained in a similar way.

Now, we add our auxiliary edges by connecting the pair of auxiliary vertices incident to each inner edge.
The auxiliary edge for type (a) inner edges is just the edge itself. We restrict our discussion to type (b) inner edges only.
It is obvious that the auxiliary edges in a convex region do not intersect other edges in other convex regions, e.g., the union of sub-cells of $t_0, t_3, t_5$ cannot intersect with each other. These auxiliary edges split the convex union of sub-cells into two convex parts, i.e., a triangle and the remaining part, which is still a convex polygon. We reassign the triangle part to its neighboring cell, which fills the concave part.
For triangle $t_2$, which has one concave and one convex power center, once the concave part is filled by its neighboring cell, it becomes convex. The auxiliary edge of $t_2$ now lies in a convex region. It cannot intersect with any other auxiliary edges. Similarly, we reassign the regions using the auxiliary edge of $t_2$. Figure~\ref{fig:appendix}(f) shows the final result of the gap decomposition, which is equivalent to the output of the algorithm described in Section~\ref{sec:extract}.

To this end, we have shown that the splitting edges do not intersect with each other, and we can conclude that our decomposition algorithm fulfills the following properties:
\begin{itemize}
\item Each gap triangle is associated with a simple convex polygon with up to six edges (three edges from three arcs and the edges or vertices connecting the arcs).
\item All these polygons are non-intersecting except along the common edges (auxiliary splitting edge).
\item The union of these polygons covers the whole gap.
\end{itemize}


\end{document}